\documentstyle[aps,12pt,epsfig]{revtex}

\parskip=\medskipamount

\def\be{\begin{equation}}
\def\ee{\end{equation}}
\def\disp{\displaystyle}
\def\foot{\footnotesize}

\def\R{{\sf I\kern-.15em R}}
\def\C{\kern.1em{\raise.47ex\hbox{$\scriptscriptstyle |$}}
             \kern-.40em{\sf C}}
\def\Z{{\sf Z\kern-.45em Z}}

\def\ze{\zeta}
\def\al{\alpha}

\begin{document}

\title{Multifractality of Entangled  Random Walks and Nonuniform Hyperbolic Spaces}
\author{Raphael Voituriez$^{1}$ and Sergei Nechaev$^{1,2}$}
\address{$^{1}$Laboratoire de Physique Th\'eorique et Mod\`eles Statistiques, Universit\'e Paris Sud, \\ 91405 Orsay Cedex, France}
\address{$^{2}$L D Landau Institute for Theoretical Physics, 117940,
Moscow, Russia}
\maketitle
\bigskip

\begin{abstract}  Multifractal properties of the distribution of 
topological invariants for a model of trajectories randomly entangled with a
nonsymmetric lattice of obstacles are investigated. Using the equivalence of the model to  random walks on a locally nonsymmetric tree, statistical properties of topological invariants, such as drift and return probabilities,  have been studied by means of a renormalization group (RG)
technique. The comparison of the analytical RG--results with  numerical
simulations as well as with the rigorous results of P.Gerl and W.Woess
\cite{woess} demonstrates clearly the validity of our approach.  It is shown explicitly by
direct counting for the discrete version of the model and by conformal
methods for the continuous version that multifractality occurs when local uniformity of the phase space (which has an exponentially large number of states) has been broken.
\end{abstract}   \bigskip

\section{Introduction} \label{sect:1} 
The phenomenon of multifractality consists in a scale dependence of critical
exponents. It has been widely discussed in the literature for a wide range of issues, such as statistics of strange sets \cite{procacc},
diffusion limited aggregation \cite{dla}, wavelet transforms \cite{holsch},
conformal invariance \cite{dupl} or  statistical properties of critical wave
functions of massless Dirac fermions in a random magnetic field
\cite{kogan,chamon,castillo}.

The aim of our work is not only to describe a new model
possessing multiscaling dependence, but also to show that the phenomenon of multifractality is related to local nonuniformity of the exponentially growing ("hyperbolic") underlying
"target" phase space, through an example of entangled random walks distribution in homotopy classes. Indeed, to our knowledge, almost  all examples of multifractal
behavior for physical \cite{kogan,chamon,castillo} or more abstract
\cite{procacc,derrida} systems share one common feature---all target phase
spaces have a noncommutative structure and are locally nonuniform.

We believe that multiscaling is a much more generic physical phenomenon
compared with uniform scaling, appearing when the phase space of a
system possesses a hyperbolic structure with local symmetry breaking. Such
perturbation of local symmetry could be either regular or random---from our point of view the
details of the origin of local nonuniformity play a less significant role.

We discuss below the basic features of multifractality in a locally
nonuniform regular hyperbolic phase space. We show in particular  that a
multifractal behavior is encountered in statistical topology in the case of entangled (or knotted) random walks distribution in topological classes.

The paper is organized as follows. In Section \ref{sect:2} we consider a 2D
$N$--step random walk in a nonsymmetric array of topological obstacles and investigate
the multiscaling properties of the ``target'' phase space for a set of
specific topological invariants---the ``primitive paths''. The renormalization
group computations of mean length of the primitive path, as well as   return
probabilities to the unentangled topological state are developed in Section \ref{sect:3}. Section \ref{sect:4} is devoted to the application of conformal methods to a  geometrical analysis of multifractality in  locally nonuniform
hyperbolic spaces.

\section{Multifractality of topological invariants for random
entanglements in a lattice of obstacles} \label{sect:2}

The concept of multifractality has been formulated and clearly explained in
the paper \cite{procacc}. We begin by recalling the basic definitions of R\'enyi spectrum, which will be used in the following.

Let $\nu(C_i)$ be an  abstract invariant distribution characterizing the
probability of a dynamical system to stay in a basin of attraction of some
stable configuration $C_i$ $(i=1,2,\ldots,{\cal N})$. Taking a uniform grid
parameterized by  "balls" of size $l$, we define the family of fractal
dimensions $D_q$:
\be \label{1:Dq}
D_q=\frac{1}{q-1}\lim_{l\to 0}\frac{\disp \ln \sum_{i=1}^{\cal N}
\nu^q(C_i)}{\ln l}
\ee
As $q$ is varied, different subsets of $\nu^q$ associated with different values
of $q$ become dominant.

Let us define the scaling exponent $\alpha$ as follows
$$
\nu^q(C_i) \sim l^{\alpha\, q}
$$
where $\alpha$ can take different values corresponding to different regions of
the measure which become dominant in Eq.(\ref{1:Dq}). In particular, it is
natural to suggest that $\sum_{i=1}^{\cal N} \nu^q(C_i)$ can be rewritten as follows:
$$
\sum_{i=1}^{\cal N} \nu^q(C_i)=
\left.\left[\int d\alpha' \rho(\alpha')l^{-f(\alpha')} l^{\alpha'\, q}\right]
\right|_{l\to 0}
$$
where $\rho(\alpha)$ is the probability to have the value $\alpha$ lying in
 a small "window" $[\alpha', \,\alpha'+\Delta\alpha']$ and $f(\alpha)$ is
a continuous function which has sense of  fractal dimension of the subset
characterized by the value $\alpha$.

Supposing $\rho(\alpha)>0$ one can approximately evaluate the last expression
via the saddle--point method. Thus, one gets (see, for example, 
\cite{procacc}):
$$
\begin{array}{l}
\disp \frac{d}{d\alpha}f(\alpha)=q \medskip \\
\disp \frac{d^2}{d\alpha^2}f(\alpha)<0
\end{array}
$$
what together with (\ref{1:Dq}) leads to the following equations
\be \label{1:Dq2}
\begin{array}{l}
\disp \tau(q)=q\alpha(q)-f[\alpha(q)] \medskip \\
\disp \alpha(q)=\frac{d}{\disp dq}\tau(q)
\end{array}
\ee
where $\tau(q)=(q-1)D_q$. Hence, the exponents $\tau(q)$ and $f[\alpha(q)]$
are related via Legendre transform. For further details and more advanced
mathematical analysis,  the reader is refered to \cite{pw}.

\subsection{2D topological systems and their relation to hyperbolic
geometry} \label{1:sect:2.1}

Topological constraints essentially modify physical properties of
the broad class of statistical systems composed of chain--like objects. It should be stressed that topological problems
are widely investigated in connection with quantum field and string
theories, 2D gravitation, statistics of vortices in superconductors,
quantum Hall effect, thermodynamic properties of entangled polymers etc.
Modern methods of theoretical physics allow us to describe rather
comprehensively  the effects of nonabelian statistics on the physical behavior
of some systems. However the following
question remains still  obscure: what are the fractal (and as it is
shown below, multifractal) properties of the distribution function of
topological invariants, characterizing the homotopy states of a
statistical system with topological constraints? We investigate this problem
in the framework of the model "Random Walk in an Array of Obstacles" (RWAO).

The RWAO--model can be regarded as physically clear  and as a very representative
image for systems of fluctuating chain--like objects with a full range
of nonabelian topological properties. This model is formulated as follows: suppose that  a  random walk of $N$ steps of length $a$ takes place on a  plane
between obstacles which form a simple 2D rectangular lattice with 
unit cell of size $c_x \times c_y$. We assume that the random
walk cannot cross ("pass through") any obstacles.
\begin{figure}
\centerline{\epsfig{file=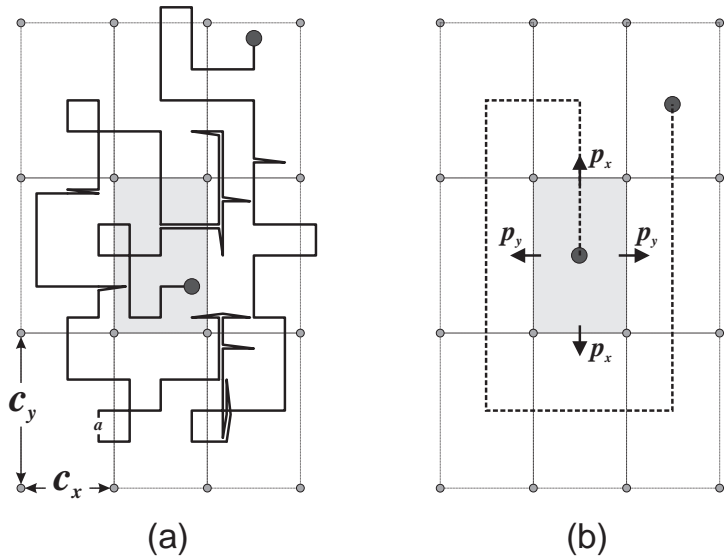,width=10cm}}
\caption{Random walk in the two-dimensional rectangular lattice of obstacles.}
\label{1:fig:1}
\end{figure}

It is convenient to begin with the lattice realization of the RWAO--model.
In this case the random path can be represented as a $N$--step random walk
in a square lattice of size $a\times a$ ($a\le c_y\le
c_x$)---see fig.\ref{1:fig:1}.

\begin{figure}
\centerline{\epsfig{file=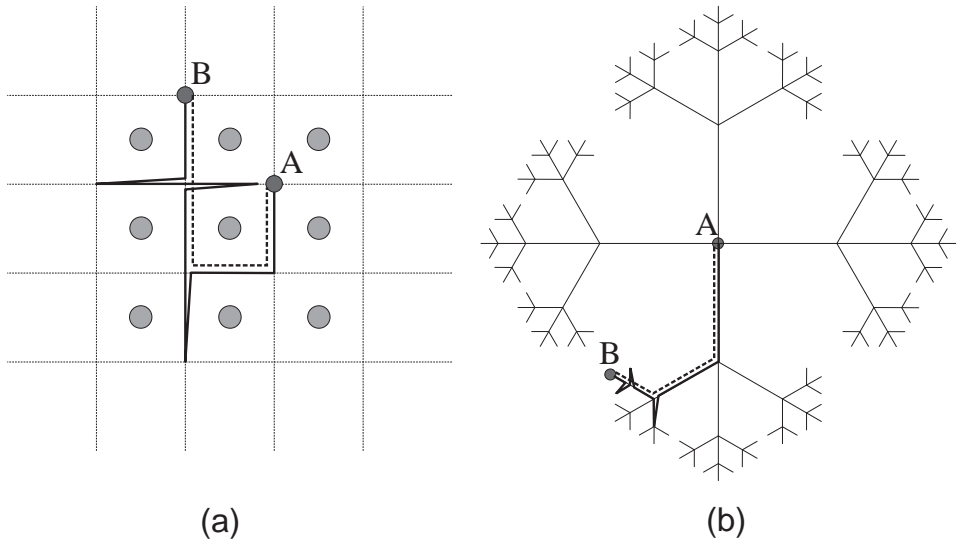,width=13cm}}
\caption{Random path for $a=c_x=c_y$: (a) in the 2D lattice of
obstacles; (b) in the covering space (on the Cayley tree).}
\label{1:fig:2}
\end{figure}

It had been shown previously (see, for example \cite{ne_kh_sem,khter}) that for
$a=c_x=c_y$ a lattice random walk in the presence of a  regular array of obstacles (punctures) on the
dual lattice ${\Z}^2$ is topologicaly equivalent to a  free random walk on a graph---a
Cayley tree with branching number $z=4$ (see Fig.\ref{1:fig:2}). An
outline of the derivation of this result is as follows. The different topological states of our problem coincide with the elements of the homotopy group of the multi-punctured plane, which is the free group $\Gamma_{\infty}$ generated by  a countable set of elements. The translational invariance allows to consider a local basis and therefore to study the factored group $\Gamma_{\infty}/{\Z}^2=\Gamma_{z/2}$, where $\Gamma_{z/2}$ is a free group with $z/2$ generators whose Cayley graph is precisely a $z$--branching tree.


The relation between  Cayley trees and hyperbolic geometry is
discussed in details in Section \ref{sect:3}. Intuitively 
such a relation could be understood as follows. The Cayley tree can be
isometrically embedded in the hyperbolic plane ${\cal H}$ (surface of
constant negative curvature). The group $\Gamma_{z/2}$ is one of the discrete
subgroups of the group of motion of the hyperbolic plane ${\cal H}=SL(2,{\R})/SO(2)$, therefore the Cayley tree can be considered as a particular discrete realization of the hyperbolic plane.

Returning to the RWAO--model, we conclude that each trajectory in the lattice
of obstacles can be lifted to a path in the ``universal covering space'' i.e to
a path on the $z$--branching Cayley tree. The geodesic on the Cayley graph, i.e the shortest trajectory along the graph which connects ends of the path, plays the
role of a complete topological invariant for the original trajectory in the
lattice of obstacles. For example, the random walk in the lattice of obstacles
is closed and contractible to a point (i.e. is not entangled with
the array of obstacles) if and only if the geodesic length between the ends of
the trajectory on the Cayley graph is zero. Hence, this geodesic length can be regarded as a topological invariant, which preserves the main  nonabelian features of the considered problem.

We would like  to stress two facts concerning our model: (i) The exact
configuration of a geodesic is a complete topological invariant, while its
length $k$ is only a partial topological invariant  (except the case $k=0$); (ii) Geodesics have
a clear geometrical interpretation, having sense of a bar (or "primitive") path
which remains after deleting all even times folded parts of a  random trajectory
in the lattice of obstacles. The concept of  "primitive path" has been repeatedly
used in statistical physics of polymers, leading to a successful classification of the topological states of chain--like molecules in various topological problems
\cite{ne_kh_sem,khter,helf_rub}.

Even if  many aspects of statistics of random walks in fixed lattices of
obstacles have been  well understood (see, for example \cite{nechaev} and
references therein), the set of problems dealing with the investigation of
fractal properties of the distribution of topological invariants  in the
RWAO--model are practically out of discussion. Thus we devote the next
Section to the study of fractal and multifractal structures of the
measure on the set  of primitive paths in the RWAO--model for $a\ll c_y<c_x$.

\subsection{Multifractality of the measure on the set of primitive
paths on a  nonsymmetric Cayley tree} \label{1:sect:2.2}

The classification of different topological states of a $N$--step random walk
in a rectangular lattice of obstacles in the case $a\ll c_y<c_x$ turns out to be
a more difficult and more rich problem than in the case $a=c_y=c_x$
discussed above. However, after a proper rescaling, the
mapping of a random walk in the rectangular array of obstacles to a
random walk on a Cayley tree can be explored again. To proceed we should
solve two auxiliary problems. First of all we consider a  random walk inside the
elementary rectangular cell of the lattice of obstacles. Let us compute:

(i) The "waiting time", i.e the average number of steps
$\left<t\right>$ which a $t$--step random walk spends within the rectangle of
size $c_x\times c_y$;

(ii) The ratio of the "escape probabilities" $p_x$ and $p_y$
through the corresponding sides $c_x$ and $c_y$ for a random walk staying
till time $t$ within the elementary cell.

The desired quantities can be easily computed from the distribution
function $P(x_0,y_0,x,y,t)$ which gives the probability to find the
$t$--step random walk with initial $(x_0,y_0)$ and final $(x,y)$ points
within the rectangle of size $c_x\times c_y$. The function $P(x,y,t)$ in the
continuous approximation ($a\to 0;\, t\to\infty;\, at={\rm const}$) is the
solution of the following boundary problem
\be \label{1:2}
\left\{\begin{array}{l}
\disp \frac{\partial}{\partial t}P(x,y,t) =
\frac{a^2}{4}\left(\frac{\partial^2}{\partial x^2}+
\frac{\partial^2}{\partial y^2}\right)P(x,y,t) \medskip \\
P(0,y,t)=P(c_x,y,t)=P(x,0,t)=P(x,c_y,t)=0 \medskip \\
P(x,y,0)=\delta(x_0,y_0)
\end{array}
\right.
\ee
where $a$ is the length of the effective step of the random walk and the value
$\frac{a^2}{4}$ has sense of a  diffusion constant.

The solution of Eqs.(\ref{1:2}) reads
\be \label{1:3}
P(x_0,y_0,x,y,t)=\frac{4}{c_x\;c_y} \sum_{m_x=1}^{\infty}\sum_{m_y=1}^{\infty}
e^{-\frac{\pi^2 a^2}{4}\left(\frac{m_x^2}{c_x^2}+\frac{m_y^2}{c_y^2}
\right)t}\; \sin\frac{\pi m_x x_0}{c_x}\;\sin\frac{\pi m_y y_0}{c_y}\;
\sin\frac{\pi m_x x}{c_x}\;\sin\frac{\pi m_y y}{c_y}
\ee

The "waiting time" $\left<t\right>$ can be written now as follows
\be \label{1:3a}
\left<t\right>=\frac{1}{c_x\;c_y}\int_0^{c_x} dx_0 \int_0^{c_y} dy_0
\int_0^{c_x} dx \int_0^{c_y} dy \int_0^{\infty}dt\; P(x_0,y_0,x,y,t)
\ee
while the ratio $p_x/p_y$ can be computed straightforwardly via the
relation:
\be \label{1:3b}
\frac{p_x}{p_y}=\frac{\disp \left.\int_0^{c_x} dx_0 \int_0^{c_y} dy_0
\int_0^{c_x} dx\; P(x_0,y_0,x,y,t)\, \right|_{y=\{a,c_y-a\}}}
{\disp \left.\int_0^{c_x} dx_0 \int_0^{c_y} dy_0 \int_0^{c_y} dy\;
P(x_0,y_0,x,y,t)\, \right|_{x=\{a,c_x-a\}}}
\ee

In the "ground state dominance" approximation we truncate the sum (\ref{1:3})
at $m_x=m_y=1$ and get  the following approximate expressions:
\be \label{1:4a}
\left<t\right>=\frac{4^4\, c_x^2\,c_y^2}{\pi^6\, a^2\, (c_x^2+c_y^2)};\qquad
\frac{p_x}{p_y}=\frac{c_x^2}{c_y^2}
\ee
In the symmetric case ($c_x=c_y\equiv c$) Eq.(\ref{1:4a}) gives 
$\left<t\right>=\frac{2^7\,c^2}{\pi^6\,a^2}$ and $p_x/p_y=1$,  as it should be
for the square lattice of obstacles.

Now the distribution function of the primitive paths for the RWAO model can be obtained
via lifting this topological problem to the problem of {\it directed} random
walks\footnote{Recall that by definition the primitive path is the geodesic
distance and therefore cannot have two successive opposite steps.} on the
4--branching Cayley tree, where the random walk on the Cayley tree is
defined as follows:

(a) The total number of steps $\tilde{N}$ on the Cayley tree is
$$
\tilde{N}=\frac{N}{\left<t\right>}=
\frac{\pi^6}{4^4}\,\frac{N a^2(c_x^2+c_y^2)}{c_x^2\,c_y^2}
$$
(the value $\left<t\right>$ has been computed in (\ref{1:4a})).

\begin{figure}
\begin{center}
\unitlength=1.00mm
\special{em:linewidth 0.6pt}
\linethickness{0.6pt}
\begin{picture}(153.33,55.00)
\put(80.00,55.00){\line(-1,-1){20.00}}
\put(80.00,55.00){\line(1,-1){20.00}}
\put(80.00,55.00){\line(3,-1){60.00}}
\put(80.00,55.00){\line(-3,-1){60.00}}
\put(20.00,35.00){\line(-2,-3){10.00}}
\put(20.00,35.00){\line(0,-1){15.00}}
\put(20.00,35.00){\line(2,-3){10.00}}
\put(60.00,35.00){\line(-2,-3){10.00}}
\put(60.00,35.00){\line(0,-1){15.00}}
\put(60.00,35.00){\line(2,-3){10.00}}
\put(100.00,35.00){\line(-2,-3){10.00}}
\put(100.00,35.00){\line(0,-1){15.00}}
\put(100.00,35.00){\line(2,-3){10.00}}
\put(140.00,35.00){\line(-2,-3){10.00}}
\put(140.00,35.00){\line(0,-1){15.00}}
\put(140.00,35.00){\line(2,-3){10.00}}
\put(50.00,20.00){\line(-1,-3){3.33}}
\put(50.00,20.00){\line(0,-1){10.00}}
\put(50.00,20.00){\line(1,-3){3.33}}
\put(60.00,20.00){\line(-1,-3){3.33}}
\put(60.00,20.00){\line(0,-1){10.00}}
\put(60.00,20.00){\line(1,-3){3.33}}
\put(70.00,20.00){\line(-1,-3){3.33}}
\put(70.00,20.00){\line(0,-1){10.00}}
\put(70.00,20.00){\line(1,-3){3.33}}
\put(10.00,20.00){\line(-1,-3){3.33}}
\put(10.00,20.00){\line(0,-1){10.00}}
\put(10.00,20.00){\line(1,-3){3.33}}
\put(20.00,20.00){\line(-1,-3){3.33}}
\put(20.00,20.00){\line(0,-1){10.00}}
\put(20.00,20.00){\line(1,-3){3.33}}
\put(30.00,20.00){\line(-1,-3){3.33}}
\put(30.00,20.00){\line(0,-1){10.00}}
\put(30.00,20.00){\line(1,-3){3.33}}
\put(130.00,20.00){\line(-1,-3){3.33}}
\put(130.00,20.00){\line(0,-1){10.00}}
\put(130.00,20.00){\line(1,-3){3.33}}
\put(140.00,20.00){\line(-1,-3){3.33}}
\put(140.00,20.00){\line(0,-1){10.00}}
\put(140.00,20.00){\line(1,-3){3.33}}
\put(150.00,20.00){\line(-1,-3){3.33}}
\put(150.00,20.00){\line(0,-1){10.00}}
\put(150.00,20.00){\line(1,-3){3.33}}
\put(90.00,20.00){\line(-1,-3){3.33}}
\put(90.00,20.00){\line(0,-1){10.00}}
\put(90.00,20.00){\line(1,-3){3.33}}
\put(100.00,20.00){\line(-1,-3){3.33}}
\put(100.00,20.00){\line(0,-1){10.00}}
\put(100.00,20.00){\line(1,-3){3.33}}
\put(110.00,20.00){\line(-1,-3){3.33}}
\put(110.00,20.00){\line(0,-1){10.00}}
\put(110.00,20.00){\line(1,-3){3.33}}
\put(38.00,44.00){\makebox(0,0)[cc]{$1$}}
\put(74.00,44.00){\makebox(0,0)[cc]{$\beta$}}
\put(86.00,44.00){\makebox(0,0)[cc]{$1$}}
\put(123.00,44.00){\makebox(0,0)[cc]{$\beta$}}
\put(22.00,24.00){\makebox(0,0)[cc]{$1$}}
\put(10.00,24.00){\makebox(0,0)[cc]{$\beta$}}
\put(31.00,24.00){\makebox(0,0)[cc]{$\beta$}}
\put(102.00,24.00){\makebox(0,0)[cc]{$1$}}
\put(90.00,24.00){\makebox(0,0)[cc]{$\beta$}}
\put(111.00,24.00){\makebox(0,0)[cc]{$\beta$}}
\put(62.00,24.00){\makebox(0,0)[cc]{$\beta$}}
\put(50.00,24.00){\makebox(0,0)[cc]{$1$}}
\put(71.00,24.00){\makebox(0,0)[cc]{$1$}}
\put(142.00,24.00){\makebox(0,0)[cc]{$\beta$}}
\put(130.00,24.00){\makebox(0,0)[cc]{$1$}}
\put(151.00,24.00){\makebox(0,0)[cc]{$1$}}
\put(7.00,7.00){\makebox(0,0)[cc]{{\foot $1$}}}
\put(10.00,7.00){\makebox(0,0)[cc]{{\foot $\beta$}}}
\put(13.00,7.00){\makebox(0,0)[cc]{{\foot $1$}}}
\put(27.00,7.00){\makebox(0,0)[cc]{{\foot $1$}}}
\put(30.00,7.00){\makebox(0,0)[cc]{{\foot $\beta$}}}
\put(33.00,7.00){\makebox(0,0)[cc]{{\foot $1$}}}
\put(17.00,7.00){\makebox(0,0)[cc]{{\foot $\beta$}}}
\put(20.00,7.00){\makebox(0,0)[cc]{{\foot $1$}}}
\put(23.00,7.00){\makebox(0,0)[cc]{{\foot $\beta$}}}
\put(57.00,7.00){\makebox(0,0)[cc]{{\foot $1$}}}
\put(60.00,7.00){\makebox(0,0)[cc]{{\foot $\beta$}}}
\put(63.00,7.00){\makebox(0,0)[cc]{{\foot $1$}}}
\put(47.00,7.00){\makebox(0,0)[cc]{{\foot $\beta$}}}
\put(50.00,7.00){\makebox(0,0)[cc]{{\foot $1$}}}
\put(53.00,7.00){\makebox(0,0)[cc]{{\foot $\beta$}}}
\put(67.00,7.00){\makebox(0,0)[cc]{{\foot $\beta$}}}
\put(70.00,7.00){\makebox(0,0)[cc]{{\foot $1$}}}
\put(73.00,7.00){\makebox(0,0)[cc]{{\foot $\beta$}}}
\put(87.00,7.00){\makebox(0,0)[cc]{{\foot $1$}}}
\put(90.00,7.00){\makebox(0,0)[cc]{{\foot $\beta$}}}
\put(93.00,7.00){\makebox(0,0)[cc]{{\foot $1$}}}
\put(107.00,7.00){\makebox(0,0)[cc]{{\foot $1$}}}
\put(110.00,7.00){\makebox(0,0)[cc]{{\foot $\beta$}}}
\put(113.00,7.00){\makebox(0,0)[cc]{{\foot $1$}}}
\put(97.00,7.00){\makebox(0,0)[cc]{{\foot $\beta$}}}
\put(100.00,7.00){\makebox(0,0)[cc]{{\foot $1$}}}
\put(103.00,7.00){\makebox(0,0)[cc]{{\foot $\beta$}}}
\put(137.00,7.00){\makebox(0,0)[cc]{{\foot $1$}}}
\put(140.00,7.00){\makebox(0,0)[cc]{{\foot $\beta$}}}
\put(143.00,7.00){\makebox(0,0)[cc]{{\foot $1$}}}
\put(127.00,7.00){\makebox(0,0)[cc]{{\foot $\beta$}}}
\put(130.00,7.00){\makebox(0,0)[cc]{{\foot $1$}}}
\put(133.00,7.00){\makebox(0,0)[cc]{{\foot $\beta$}}}
\put(147.00,7.00){\makebox(0,0)[cc]{{\foot $\beta$}}}
\put(150.00,7.00){\makebox(0,0)[cc]{{\foot $1$}}}
\put(153.00,7.00){\makebox(0,0)[cc]{{\foot $\beta$}}}
\end{picture}
\end{center}
\caption{4--branching Cayley tree with different transition probabilities
along branches.}
\label{fig:4_cayley}
\end{figure}
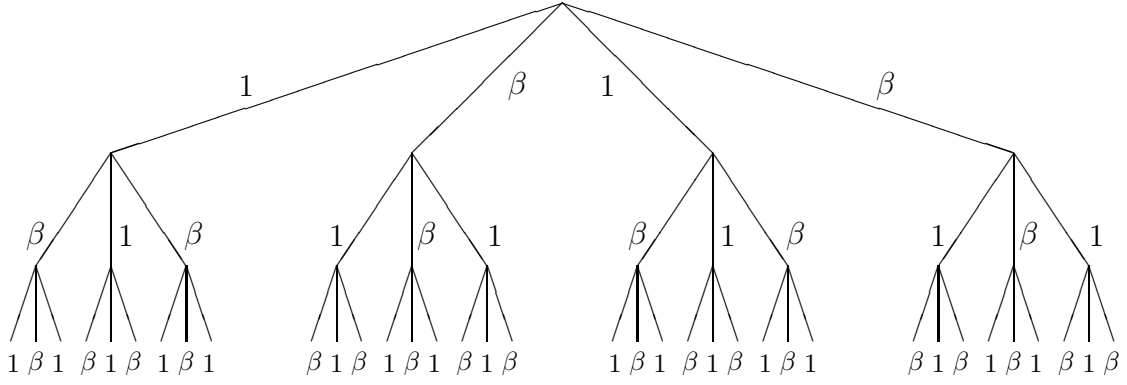

(b) The distance (or "level" $k$) on the Cayley tree is defined as the number of steps of the shortest path between two points on the tree. Each
vertex of the Cayley tree has 4 branches; the steps along two of them carry a Boltzmann weight $1$, while the steps along the two remaining ones carry a Boltzmann weight $\beta$ as it is shown in Fig.\ref{fig:4_cayley}. 
The value of $\beta$ is fixed by Eq.(\ref{1:4a}), which yields
\be \label{1:4c}
\beta=\frac{p_x}{p_y}=\frac{c_x^2}{c_y^2}
\ee

The ultrametric structure of the topological phase space, i.e. of the Cayley
tree $\gamma(\beta)$, allows us to use the results of  paper
\cite{procacc} for investigating multicritical properties of the measure of
all primitive (directed) paths of $k$ steps along the graph
$\gamma(\beta)$ with nonsymmetric weights $1$ and $\beta$ (see 
Fig.\ref{fig:4_cayley}). A rigorous  mathematical description of such weighted paths on trees (called cascades) can be found in \cite{holley}, where the authors derive multifractal spectra, but  for different distributions of weights.

We construct the partition function $\Omega(\beta,k)$ which
counts properly the weighted number of all $4\times 3^{k-1}$ different $k$--step
primitive paths on the graph $\gamma(\beta)$.

Define two partition functions $a_k$ and $b_k$ of $k$--step paths, whose last
steps carry the weights $1$ and $\beta$ correspondingly. These functions
satisfy the recursion
relations for $k\ge 1$:
\be \label{1:6}
\left\{\begin{array}{l}
a_{k+1}=a_k+2\,b_k \\
b_{k+1}=2\beta\, a_k+\beta\, b_k
\end{array}
\right. \qquad (k\ge 1)
\ee
with the following initial conditions at $k=1$:
\be \label{1:6a}
\left\{\begin{array}{l}
a_1=2 \\
b_1=2\beta
\end{array}
\right.
\ee

Combining (\ref{1:6}) and (\ref{1:6a}) we arrive at the following 2--step
recursion relation for the function $a_k$:
\be \label{1:6b}
\left\{\begin{array}{ll}
a_{k+2}=(1+\beta)\, a_{k+1}+3\beta\, a_k & \qquad(k\ge 1) \\
a_1=2 & \qquad(k=1) \\
a_2=2+4\beta & \qquad(k=2)
\end{array}
\right.
\ee
whose solution is
\be \label{1:6c}
a_k=\frac{a_2-a_1\lambda_2}{\lambda_1-\lambda_2}\;
\lambda_1^{k-1}+\frac{a_1\lambda_1-a_2}{\lambda_1-\lambda_2}\;
\lambda_2^{k-1}
\ee
where
\be \label{1:6d}
\lambda_{1,2}=\frac{1}{2}\left(1+\beta \pm\sqrt{(1+\beta)^2+12\beta}\right)
\ee

Taking into account that $b_k$ is given by the same recursion relation as
$a_k$ but with the initial values $b_1=2\beta$ and $b_2=2\beta^2+4\beta$, we 
get the following expression for the partition function $\Omega(\beta,k)=
a_k+b_k$:
\be \label{1:6sum}
\Omega(\beta,k)=
\frac{2(1+4\beta+\beta^2)-2(1+\beta)\lambda_2}{\lambda_1-\lambda_2}\;
\lambda_1^{k-1}+
\frac{2(1+\beta)\lambda_1-2(1+4\beta+\beta^2)}{\lambda_1-\lambda_2}\;
\lambda_2^{k-1}
\ee

The partition function $\Omega(\beta,k)$ contains all necessary information
about the multifractal behavior. Following Eqs.(\ref{1:Dq})--(\ref{1:Dq2}), we
associate the set of stable configurations $\{C_i\}$ with the set of
${\cal N}(k)=4\times 3^{k-1}$ vertices of level $k$.
Hence, we define
\be \label{1:set}
\sum_{i=1}^{\cal N} \nu^q(C_i)=\frac{\Omega(\beta^q, k)}{\Omega^q(\beta, k)}
\ee

\begin{figure}[ht]
\begin{center}
\epsfig{file=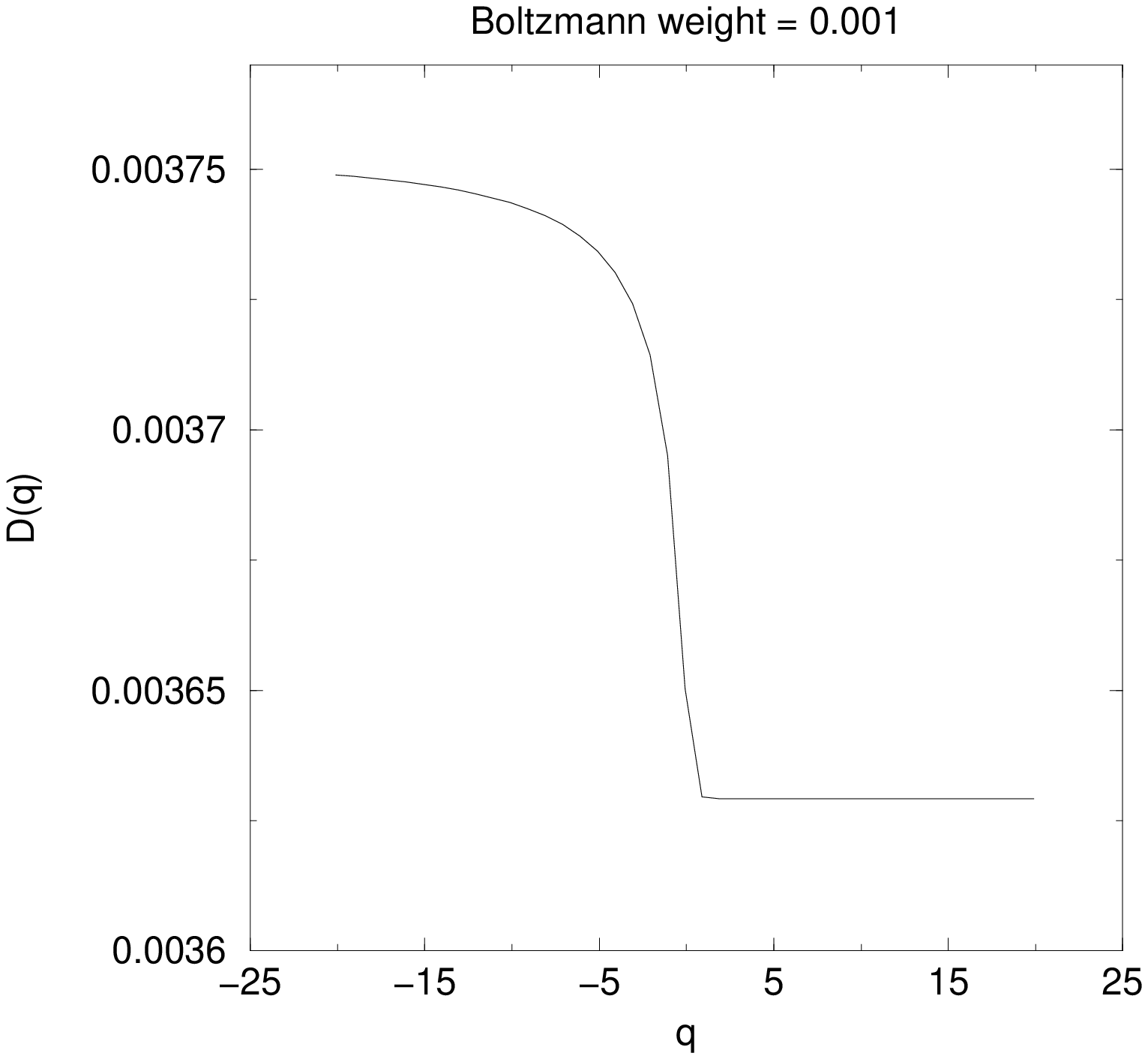,width=7cm}
\epsfig{file=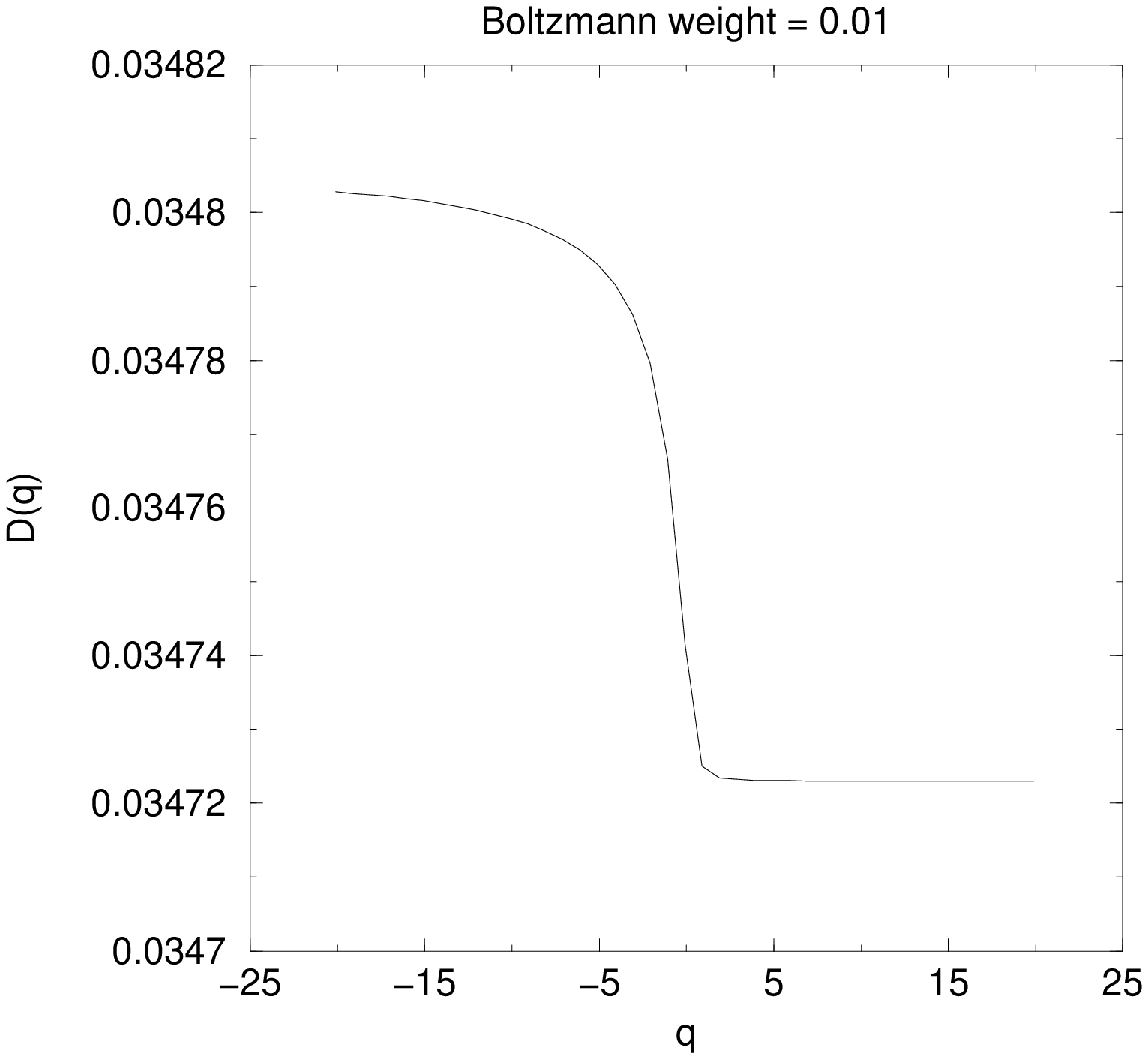,width=7cm} \\
\epsfig{file=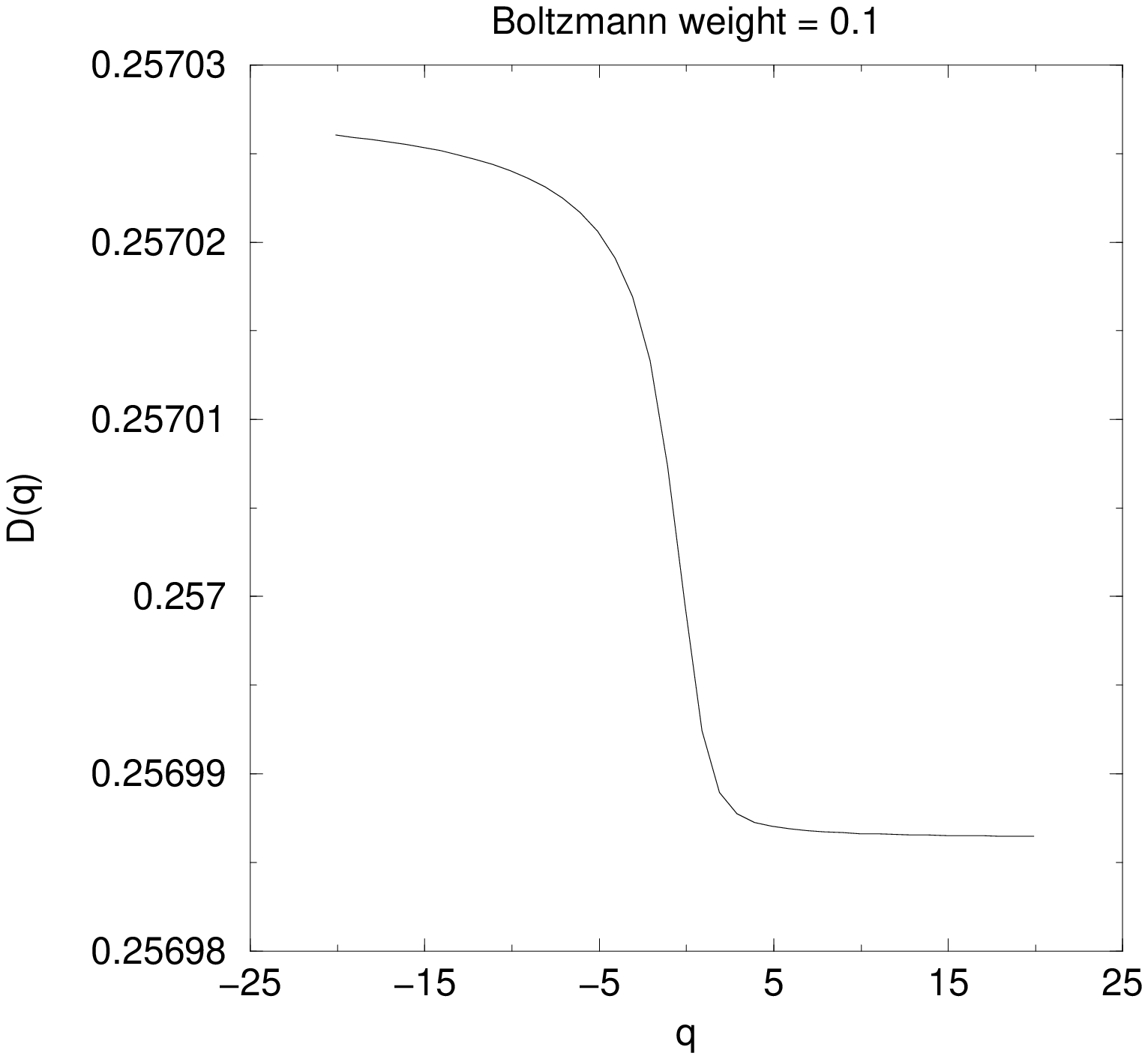,width=7cm}
\epsfig{file=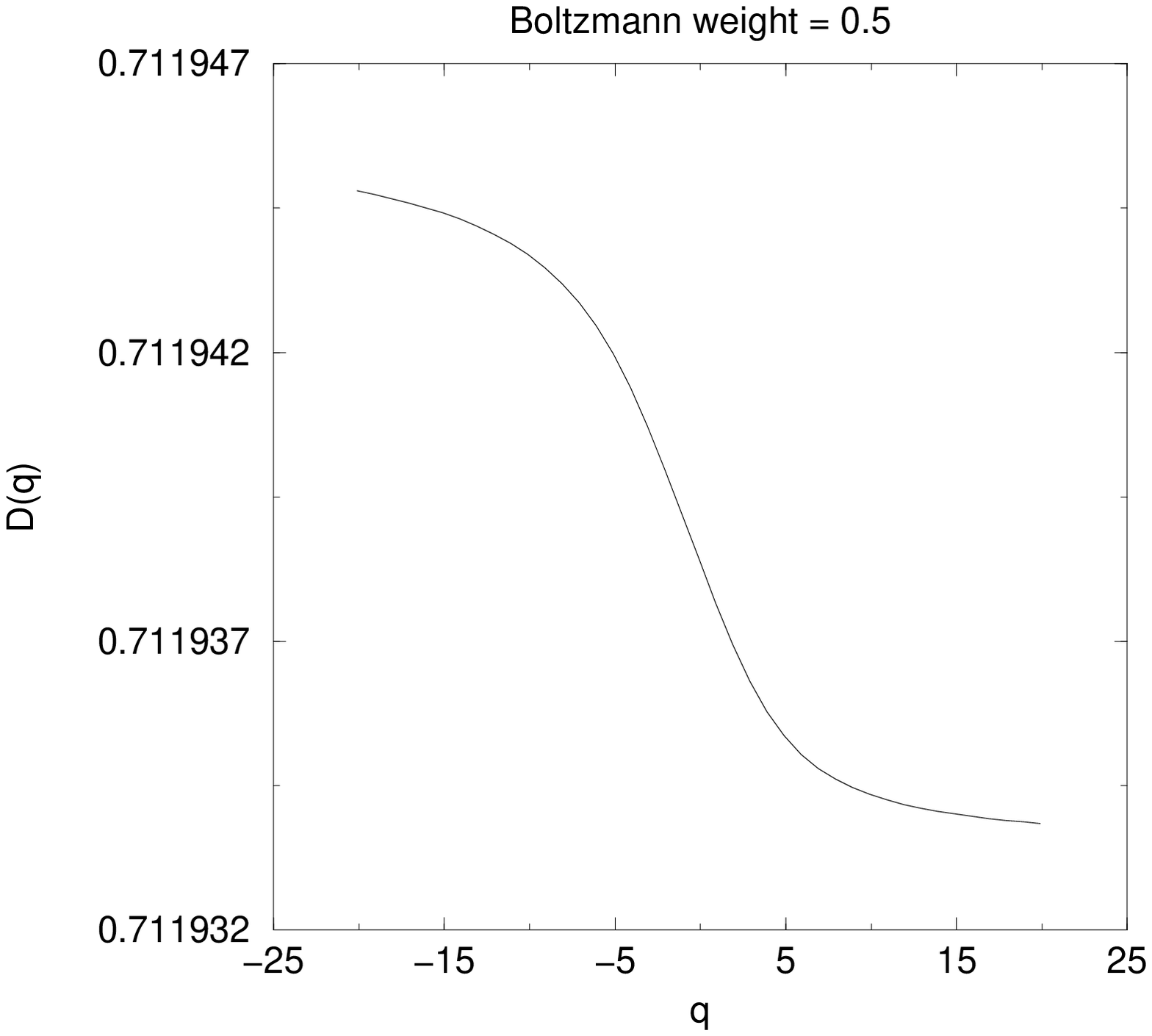,width=7cm}
\end{center}
\caption{Dependence $D_q(q)$ for different values of Boltzmann weight
$\beta$.}
\label{1:fig:4}
\end{figure}

Taking into account that the uniform grid has resolution $l(k)=1/{\cal N}(k)$
for $k\ge 1$ and using Eq.(\ref{1:Dq}), we obtain
\be \label{1:7}
\tau(q)=\lim_{k\to \infty}\frac{\ln\Omega(\beta^q,k)-
q\ln\Omega(\beta,k)}{\ln l(k)}
\ee
which allows to determine the generalized Hausdorff dimension $D_q$ via the
relation
\be \label{1:7a}
D_q=\tau(q)/(q-1)
\ee

The corresponding plots of the functions $D_q(q)$ for different values of $\beta=\{0.001;\,0.01;\,0.1;\,0.5\}$ are shown in
Fig.\ref{1:fig:4} (the numerical computations of Eqs.(\ref{1:7})--(\ref{1:7a})
are carried out for $k=100\,000$). The fact that $D_q(q)$ depends on $q$ clearly demonstrate the multifractal behavior.

\section{Random walk on a nonsymmetric Cayley tree} \label{sect:3}
\subsection{Master equation}

Consider a  random walk on a 4-branching Cayley tree and investigate the distribution $P(k,\tilde{N})$ giving the probability for a  $\tilde{N}$--step random walk
starting at the origin of the tree to have a  primitive (shortest) path
between ends of length $k$. The random
walk is defined as follows: at each vertex of the Cayley tree the probability
of a step along two of the branches is $p_{x}$,  and is $p_{y}$ along the two others;
$p_{x}$ and $p_{y}$ satisfy the conservation condition $2p_{x}+2p_{y}=1$. Using
Eq.(\ref{1:4c}), the following expressions hold:
\be \label{2:pxpy}
\left\{\begin{array}{l}
\disp p_x=\frac{\beta}{2(1+\beta)} \medskip \\
\disp p_y=\frac{1}{2(1+\beta)}
\end{array}\right.
\ee

The symmetric case $\beta=1$ (which gives $p_{x}=p_{y}=1/4$) has already  been
studied and an exact expression for $P(k,\tilde{N})$ has been derived in
\cite{kesten}. The rigorous  mathematical description  of random walks on graphs can be found in \cite{woess2}. Importance of spherical symmetry (i-e the fact that all vertices of a given level are strictly equivalent) is discussed in \cite{lyons}. Another example of nonsymmetric model on a tree (case of randomly distributed transition probabilities, the so-called RWRE model) is described in \cite{pem}. To our  knowledge, the solution for  the nonsymmetric random walk which we defined above is 
known only for $k$ fixed and $\tilde{N}\gg1$ \cite{woess}. Here we consider the case  $k\gg 1,\;\tilde{N}\gg 1$, and in particular we study the distribution in the neighborhood of the maximum.  Breaking
the symmetry by taking $\beta\neq 1$ affects  strongly the structure of the
problem, since then  the phase space becomes locally  nonuniform: we have now vertices  of
two different kinds, $x$ and $y$, depending on  whether the step toward the
root of the Cayley tree occurs with probability $p_{x}$  or $p_{y}$. In order to 
obtain a master equation for $P(k,\tilde{N})$, we  introduce the new variables
$L_{x}(k,\tilde{N})$ and $L_{y}(k,\tilde{N})$, which define the  probabilities
to be at the level $k$ in a vertex $x$ or $y$ after $\tilde{N}$ steps. We recursively define the same way the probabilities $L_{a_{1}...a_{n}}(k,\tilde{N})$, 
($a_{i}=\{x,y\}$) to be at level $k$ in a vertex such  that the sequence  of
vertices toward the root of the tree is $a_{1}...a_{n}$. One can see that the recursion 
depends on the total "history" till the root point, what makes the problem  
nonlocal. The master equation for the distribution function $P(k,\tilde{N})$ 
\be \label{2:p11}
\begin{array}{lll} P(k,\tilde{N}+1) & = &
(2p_{x}+p_{y})L_{y}(k-1,\tilde{N})+(2p_{y}+p_{x})L_{x}(k-1,\tilde{N})+ \\ & &
p_{y}L_{y}(k+1,\tilde{N})+p_{x}L_{x}(k+1,\tilde{N}) 
\end{array} 
\ee 
is coupled to the hierarchical set of functions $\{L_x,L_y;L_{xx},L_{xy},
L_{yx},L_{yy};...;L_{a_1...a_n}\}$ which satisfy the following recursion relation 
\be \label{2:p12} 
\begin{array}{lll}
L_{a_{1}...a_{n}}(k,\tilde{N}+1) & = & (2-\delta_{a_{1},a_{2}})
p_{a_{1}}L_{a_{2}...a_{n}}(k-1,\tilde{N})+ \\ 
& & p_{x}L_{xa_{1}...a_{n}}(k+1,\tilde{N})+p_{y}L_{ya_{1}...a_{n}}(k+1,\tilde{N}) 
\end{array} 
\ee
where $a_{1}...a_{n}$ cover all sequences of any lengths ($\le k$) in $x$ and
$y$. In order to close this infinite system at an arbitrary order $n_0$ we make the 
following assumption: for any $n\leq  n_0$ we have 
\be\label{2:ap}
\left.\frac{L_{a_{1}...a_{n}}(k,\tilde{N})}{P(k,\tilde{N})}\right|_{k\gg n_0\atop \tilde{N}\gg n_0}
\longrightarrow \alpha_{a_{1}...a_{n}} 
\ee 
with $\alpha_{a_{1}...a_{n}}$ constant.

Using the approximation (\ref{2:ap}) we rewrite (\ref{2:p11})--(\ref{2:p12})
for large $k$ and $\tilde{N}$ in terms of the function $P(k,\tilde{N})$ and constants
$\alpha_{a_{1}...a_{n}}$ ($0<n\leq n_{0}$). Taking into account that
$$
L_{a_{1}...a_{n}x}+L_{a_{1}...a_{n}y}=L_{a_{1}...a_{n}}
$$
we arrive at $2^{n_0-1}$ independent recursion relations for one and the same function $P(k,\tilde{N})$, with
$2^{n_0}-1$ independent unknown constants $\alpha_{a_{1}...a_{n_0}}$. In order
to make this system self--consistent, one has to identify coefficients entering in 
different equations, what yields  $2^{n_0}-2$
compatibility relations for the constants $\alpha_{a_{1}...a_{n_0}}$, and the system is still open.
This fact means that all scales are involved and the evolution of
$L_{a_{1}...a_{n}}$ depends on $L_{a_{1}...a_{n+1}}$, the evolution of
$L_{a_{1}...a_{n+1}}$ depends on $L_{a_{1}...a_{n+2}}$ and so on. At each scale
we need informations about larger scales. This kind of scaling problem naturally
suggests to use a renormalization group approach, which is developed in the next
Section.

To begin with the renormalization procedure, we need to estimate the values of 
the constants $\alpha_{a_{1}...a_{n_0}}$ for the first (i.e. the smallest) 
scale. Let us denote
$$
\left\{\begin{array}{l}
\alpha_{x}=\alpha \\ \alpha_{y}=1-\alpha \end{array}\right.
$$
and define $\alpha_{xx},\alpha_{xy},\alpha_{yy},\alpha_{yx}$ as follows:
$$
\left\{\begin{array}{l}
\alpha_{xx}=v_{x}\alpha \\ \alpha_{yy}=v_{y}(1-\alpha) \\
\alpha_{xy}=(1-v_{x})\alpha \\ \alpha_{yx}=(1-v_{y})(1-\alpha)
\end{array}\right.
$$
Now we set
\be\label{2:ap2}
p_{x}\alpha_{xa_{1}...a_{n}}+p_{y}\alpha_{ya_{1}...a_{n}}=
\Big(p_x\alpha+p_y(1-\alpha)\Big)\alpha_{a_{1}...a_{n}}
\ee
what means that we neglect the correlations between the constants
$\alpha_{a_{1}...a_{n}}$ and $\alpha_{a_{2}...a_{n}}$ at different
scales. As it is shown in the next Section, the renormalization group approach allows us to get rid of the approximation (\ref{2:ap2}).

With (\ref{2:ap2}) one can obtain the following generic master equation
\be\label{2:sys1}
P(k,\tilde{N}+1) = \frac{p_{a_{1}}\alpha_{a_{2}...a_{n}}}{\alpha_{a_{1}...a_{n}}}\,
(2-\delta_{a_{1},a_{2}})P(k-1,\tilde{N})+\Big(\alpha p_x+(1-\alpha)p_y\Big)P(k+1,\tilde{N})
\ee
where $a_{1}...a_{n}$ again cover all possible sequences in $x$ and $y$. We 
have now $2^{n_0}-1$ unknown quantities with $2^{n_0}-1$ compatibility
relations (\ref{2:sys1}), what makes the system (\ref{2:sys1}) closed.

For illustration, we derive the solution for $n_0=2$:
\be\label{2:2step}
\left\{\begin{array}{l}
\disp P(k,\tilde{N}+1)=\frac{p_{x}}{v_x}P(k-1,\tilde{N})+
\Big(\alpha p_x+(1-\alpha)p_y\Big)P(k+1,\tilde{N})\medskip \\
\disp P(k,\tilde{N}+1)=\frac{2p_{x}(1-\al)}{\al(1-v_x)}P(k-1,\tilde{N})+
\Big(\alpha p_x+(1-\alpha)p_y\Big)P(k+1,\tilde{N})\medskip \\
\disp P(k,\tilde{N}+1)=\frac{p_{y}}{v_y}P(k-1,\tilde{N})+
\Big(\alpha p_x+(1-\alpha)p_y\Big)P(k+1,\tilde{N})\medskip \\
\disp P(k,\tilde{N}+1)=\frac{2p_{y}\al}{(1-\al)(1-v_y)}P(k-1,\tilde{N})+
\Big(\alpha p_x+(1-\alpha)p_y\Big)P(k+1,\tilde{N})
\end{array}
\right.
\ee
Note that (\ref{2:2step}) displays clearly a ${\Z}_2$ symmetry: $p_x\to p_y,\;
\al\to -\al,\; v_x\to v_y$. Compatibility
conditions for system (\ref{2:2step}) read:
\be \label{2:values}
\frac{p_x}{v_x}=\frac{p_y}{v_y}=
\frac{2p_{x}(1-\al)}{\al(1-v_x)}=\frac{2p_{y}\al}{(1-\al)(1-v_y)}
\ee
which finally gives
\be\label{2:sol}
\left\{\begin{array}{l}
\disp \alpha=\frac{-1-3\beta+\sqrt{1+14\beta+\beta^2}}{2(1-\beta)} \medskip \\
\disp v_x=\frac{\alpha}{2-\alpha} \medskip \\
\disp v_y=\frac{1-\alpha}{1+\alpha}
\end{array} \right.
\ee
As it has been said above, without (\ref{2:ap2}) the system
(\ref{2:p11})--(\ref{2:p12}) is open, giving a single equation for the unknown
function $P(k,\tilde{N})$ depending on the unknown parameter $\al$:
\be\label{2:p2}
\begin{array}{lll}
P(k,\tilde{N}+1) & = & \Big((2p_{x}+p_{y})(1-\alpha)+
(2p_{y}+p_{x})\alpha\Big)P(k-1,\tilde{N})+ \medskip \\
& & \Big(p_{y}(1-\alpha)+p_{x}\alpha\Big)P(k+1,\tilde{N})
\end{array}
\ee
Eq.(\ref{2:p2}) describes a  1D diffusion process with a drift
\be\label{drift}
\frac{\langle k\rangle}{\tilde{N}}\equiv\overline{k}=2\alpha p_y+2(1-\alpha)p_x
\ee
and a dispersion
\be\label{dispersion}
\delta=\frac{\langle k-\langle k\rangle\rangle^2}{\tilde{N}}=
1-4\Big(\al p_y+(1-\al)p_x\Big)^2
\ee
which provides for $k\gg 1$ and $\tilde{N}\gg 1$ the usual Gaussian distribution with
nonzero mean (see \cite{ne_kh_sem}). The value of $\alpha$ obtained in
(\ref{2:sol}) using the approximation (\ref{2:ap2}) gives a fair estimate of  the drift  compared with the numerical
simulations, as it is shown in Fig.\ref{2:num}.

\subsection{Real space renormalization}

In order to improve the results obtained above, we recover the information
lost in the approximation (\ref{2:ap2}) and take into account 
``interactions'' between different scales. Namely, we follow the renormalization
flow of the parameter $\alpha(l)$ at a scale $l$ supposing that a new effective
step is  a composition of $2^l$ initial lattice steps. Let us define:
\begin{itemize}
\item the probability $f_a(l)$ of going forth (with respect to the location of
the root point of the Cayley tree) from a vertex of kind $a$;
\item the probability $b_a(l)$ of going back (towards the root point of the
Cayley tree) from a vertex of kind $a$;
\item the probability $\alpha(l)$ of being at a vertex of kind $x$;
\item the conditional probability $w_a(l)$ to reach  a vertex of kind $a$ starting
from a vertex of kind $a$ under the condition that the step is forth;
\item the conditional probability $v_a(l)$ to reach  a vertex of kind $a$ starting
from a vertex of kind $a$ under the condition that the step is back;

\item the effective length $d(l)$ of a composite step.
\end{itemize}

Then the drift $\overline{k}(l)$ at scale $l$ is given by (compare with \ref{drift}):
\be\label{2:kl}
\overline{k}(l)=d(l)\left[\alpha(l)\Big(f_x(l)-b_x(l)\Big)+
\Big(1-\alpha(l)\Big)\Big(f_y(l)-b_y(l)\Big)\right]
\ee

We say that the problem is  scale--independent if the flow
$\overline{k}(l)$ is invariant under the decimation procedure, i.e. with
respect to the renormalization group. We compute the flow counting the
appropriate combinations of two steps, depending on the variable considered:
\be \label{2:renorm}
\begin{array}{lll}
w_a(l+1) & = & \disp \Big(1-w_a(l)\Big)\Big(1-w_{\bar{a}}(l)\Big)+w_{a}^2(l)
\medskip \\
v_a(l+1) &= & \disp \Big(1-v_a(l)\Big)\Big(1-v_{\bar{a}}(l)\Big)+v_{a}^2(l)
\medskip \\
f_a(l+1) & = &\disp  \frac{f_a(l)\left[w_{a}(l)f_a(l)+
\Big(1-w_a(l)\Big)f_{\bar{a}}(l)\right]}{c_a(l)}
\medskip \\
b_a(l+1) & = & \disp \frac{b_a(l)\left[v_{a}(l)b_a(l)+
\Big(1-v_a(l)\Big)b_{\bar{a}}(l)\right]}{c_a(l)}
\medskip \\
d(l+1) & = &\disp  d(l)\left[\al(l)c_x(l)+\Big(1-\al(l)\Big)c_y(l)\right]
\medskip \\
\al(l+1) & = & \disp \overline{k}(l)\left[\al(l)w_x(l)+
\Big(1-\al(l)\Big)\Big(1-w_y(l)\Big)\right]+ \medskip \\
& & \disp \left(1-\overline{k}(l)\right)\left[\al(l)v_x(l)+
\Big(1-\al(l)\Big)\Big(1-v_y(l)\Big)\right]
\end{array}
\ee
where $\bar{a}=x$ when $a=y$ (and $\bar{a}=y$ when $a=x$) and the value 
$c_a(l)$ ensures the conservation condition $f_a(l+1)+b_a(l+1)=1$ because 
we do not consider the combinations of two successive steps in opposite 
directions.

The transformation of $\al$ in (\ref{2:renorm}) needs some explanations. We
consider the drift $\overline{k}(l)$ as a probability to make a (composite)
step forward. The equation for $\al$ is given by counting the different ways
of getting to  a vertex of kind $x$. One can check that $\overline{k}(l)$ given
by (\ref{2:kl}) remains invariant under such transformation, what is considered
as a verification of the scale independence (i.e. of renormalizability).

Following the standard procedure, we find the fixed points for the flow of
$\al(l)$. First of all we realize that the recursion equations for $w_a(l)$
and $v_a(l)$ can be solved independently, providing a continuous set of fixed
points: $w_{x}^0=1-w_{y}^0$ and $v_{x}^0=1-v_{y}^0$. Using the initial
conditions (\ref{2:sol}) for $v_a(l)$ and deriving straightforwardly the absent
initial conditions for $w_a(l)$, we get
\be\label{2:ic}
\left\{\begin{array}{l}
v_x(1)=v_x \medskip \\
v_y(1)=v_y \medskip \\
\disp w_x(1)=w_x=\frac{p_x}{p_x+2p_y} \medskip \\
\disp w_y(1)=w_y=\frac{p_y}{p_y+2p_x}
\end{array}\right.
\ee
(we recall that these values are obtained by taking into account the elementary
correlations for two successive steps).

With the initial conditions (\ref{2:ic}) we find the following renormalized
values $v^0$ and $w^0$ at the fixed point
\be\label{2:limv}
\left\{\begin{array}{l}
\disp v^0=v^0(\beta)=\lim_{l\to\infty}v_x(l)=1-\lim_{l\to\infty}v_y(l)=
\frac{1}{2}\left[(v_x-v_y)\prod_{n=1}^{\infty}f^{(n)}(v_x+v_y)+1\right]
\medskip \\
\disp w^0=w^0(\beta)=\lim_{l\to\infty}w_x(l)=1-\lim_{l\to\infty}w_y(l)=
\frac{1}{2}\left[(w_x-w_y)\prod_{n=1}^{\infty}f^{(n)}(w_x+w_y)+1\right]
\end{array} \right.
\ee
where $f^{(n)}(x)$ is the $n^{\rm th}$ iteration of the function
$$
f(x)=x^2-2x+2
$$

We then  obtain successively all renormalized values at the fixed point
\be\label{2:fp}
\left\{\begin{array}{l}
f_{a}^0=1 \medskip \\
b_{a}^0=0 \medskip \\
\disp d^0=\overline{k^0}=\frac{\al^0+\beta(1-\al^0)}{1+\beta} \medskip \\
\disp \al^0=\frac{v^0+\beta w^0}{1+\beta+(1-\beta)(v^0-w^0)}
\end{array}\right.
\ee
where the invariance of the drift $\overline{k}$ is taken into account:
$$
\disp \overline{k^0}=\overline{k}(1)=2p_y\al^0+2p_x(1-\al^0)=\frac{\al^0+
\beta(1-\al^0)}{1+\beta}
$$
\begin{figure}[ht]
\begin{center}
\epsfig{file=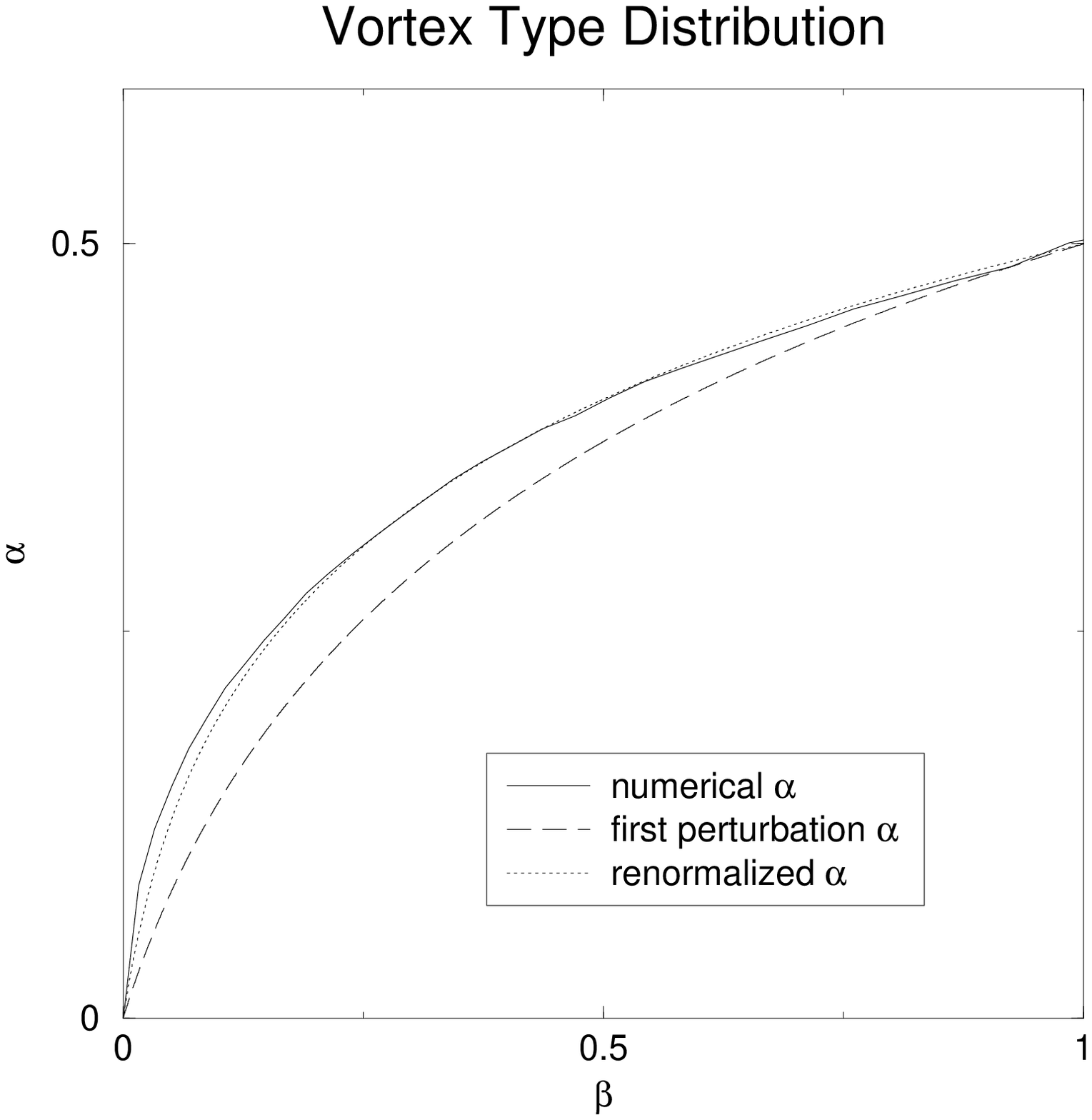,width=7cm}
\epsfig{file=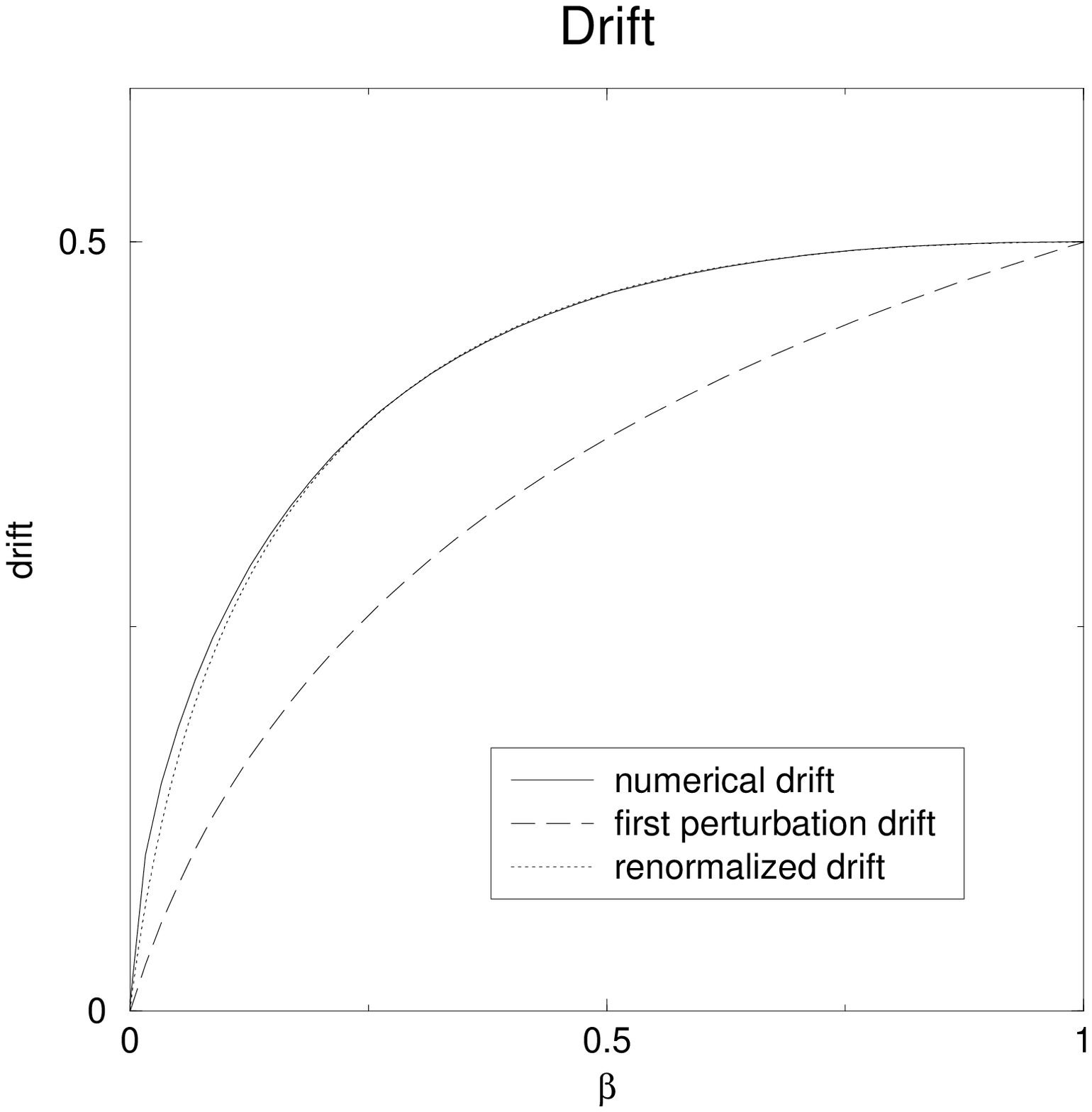,width=7cm}
\end{center}
\caption{The values of $\al$ and $\overline{k}$ compared to renormalized 
quantities and numerical simulations.}
\label{2:num}
\end{figure}

In figure \ref{2:num} we compare the theoretical results with   numerical
simulations. It is worth mentioning the efficiency of the renormalization group
method, which  yields a solution in very good agreement with numerical
simulations in a broad interval of values $\beta$.

In addition we compare our results with the exact expression obtained by P.Gerl
 and W.Woess in \cite{woess} for the probability $P(0,\tilde{N})$ to return to the 
origin after $\tilde{N}$ random steps on the nonsymmetric Cayley tree. This 
distribution function $P(0,\tilde{N})$ reads
\be \label{Woess1}
P(0,\tilde{N})\propto \mu^{\tilde{N}}\,\tilde{N}^{-3/2}
\ee
with
\be
\mu\equiv\mu(\beta)=\min\left\{-t+\sqrt{t^2+4p^{2}_x}+
\sqrt{t^2+4p^{2}_y}\;\Big|t>0\right\}
\ee

Let us assert now without justification that Eq.(\ref{2:p2}) (which is actually
written for $k\gg 1$ and $\tilde{N}\gg 1$) is valid for any values of $k$ and
$\tilde{N}$. The initial conditions for the recursion relation (\ref{2:p2}) are
as follows
\be \label{2:init}
\left\{\begin{array}{l}
P(0,\tilde{N}+1)=\Big(2\al p_x+2(1-\al) p_y\Big)P(1,\tilde{N})\\
P(k,0)=\delta_{k,0}
\end{array}\right.
\ee
One can notice that Eq.(\ref{2:p2}) completed with the conditions
(\ref{2:init}) can be viewed as a master equation for a {\it symmetric}
random walk on a Cayley tree {\it with effective branching} $z$ continuously
depending on $\beta$:
\be
z(\beta)=\frac{2}{\al p_x+(1-\al)p_y}
\ee
Hence, we conclude that our problem becomes equivalent to a symmetric
random walk on a $z(\beta)$-branching tree. For $k=0$ the solution, given
in \cite{ne_kh_sem} is
\be \label{our}
P(0,\tilde{N})\propto \left[\frac{2\sqrt{z(\beta)-1}}{z(\beta)}\right]^{\tilde{N}}\,\tilde{N}^{-3/2}
\ee
This provides the same form as the exact solution (\ref{Woess1}). It has
been checked numerically that for $\beta\in\R^+$ the discrepancy between
(\ref{Woess1}) and (\ref{our}) is as follows
$$
\frac{1}{\mu(\beta)}
\left|\frac{2\sqrt{z(\beta)-1}}{z(\beta)}-\mu(\beta)\right|<0.02
$$
Thus, we believe that our self--consistent RG--approach to statistics of
random walks on nonsymmetric trees can be extended with sufficient accuracy
to all values of $k$.

\section{Multifractality and locally nonuniform curvature of Riemann
surfaces} \label{sect:4}

We have claimed in Sections \ref{sect:1}--\ref{sect:2} that  local 
nonuniformity and the  exponentially growing  structure of the phase space of  statistical systems generates a  multiscaling behavior of the corresponding
partition functions. The aim of the present Section is to bring geometric
arguments to support  our claim by introducing a different approach of the RWAO model. The differences between  the approach considered in this Section and the one discussed in Section \ref{sect:2} are as
follows:
\begin{itemize}
\item We consider a {\it continuous} model of  random walk topologically
entangled with either a  symmetric or a nonsymmetric {\it triangular} lattice of
obstacles on the plane.
\item We pursue the goal to construct {\it explicitly} the metric structure 
of the topological phase space via conformal methods and {\it to relate directly
the nonuniform fractal relief} of the topological phase space {\it to the
multifractal properties} of the distribution function of topological invariants for
the given model.
\end{itemize}

Consider a random walk in a regular array of topological obstacles on the
plane. As in the discrete case we can split the distribution function of all
$N$--step paths with fixed positions of end points into  different topological
(homotopy) classes. We characterize each topological class by a topological invariant similar to the "primitive path" 
defined in Section \ref{sect:2}. Introducing complex coordinates $z=x+iy$ on the plane, we use conformal methods which provide an efficient tool for 
investigating multifractal properties of the distribution function of random trajectories in homotopy classes.

Let us stress that explicit expressions are constructed so far for triangular
lattices of obstacles only. That is why we replace the investigation of the
rectangular lattices discussed in Sections \ref{sect:1}--\ref{sect:2} by the
consideration of the triangular ones. Moreover, for triangular lattices a
continuous symmetry parameter (such as $\beta=c_x^2/c_y^2$ in case of
rectangular lattices) does not exist and only the triangles with angles
$(\pi/3,\pi/3,\pi/3)$, $(\pi/2,\pi/4,\pi/4)$, $(\pi/2,\pi/6,\pi/3)$ are
available---only such triangles tessellate the whole plane $z$. In spite of the
mentioned restrictions, the study of these cases enables us to figure out an origin of multifractality coming from the metric structure of the
topological phase space.

\begin{figure} 
\begin{center} 
\epsfig{file=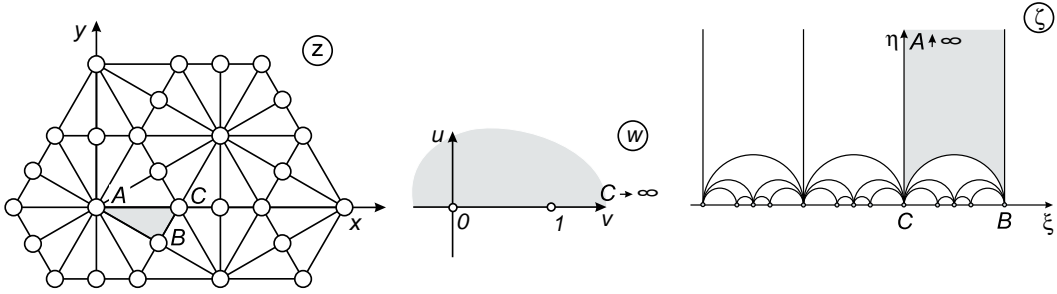,width=15cm} 
\end{center} 
\caption{Conformal mapping of the complex plane $z$ to the "lacunary" upper
half--plane ${\rm Im}\ze>0$ endowed with a Poincar\'e metric. The zero--angled
triangle ${\cal ABC}$ on $\ze$ corresponds to the triangle $ABC$ with the
angles  $(\pi/2,\pi/6,\pi/3)$ on $z$.}
\label{fig:lacun}
\end{figure}

Suppose that the topological obstacles form a periodic lattice in the $z$--plane.  Let
the fundamental domain of this lattice be the triangle $ABC$ with  angles
either $(\pi/3,\pi/3,\pi/3)$ (symmetric case) or  $(\pi/2,\pi/6,\pi/3)$ (nonsymmetric case). The conformal  mapping $z(\ze)$ establishes a
one-to-one correspondence between a given  fundamental domain $ABC$ of the
lattice of obstacles in the $z$--plane with a  zero--angled triangle ${\cal ABC}$
lying in the upper half--plane $\eta>0$  of the plane $\ze=\xi+i\eta$, and
having corners on the real axis $\eta=0$. To  avoid possible misunderstandings let
us point out that such transform is  conformal everywhere except at corner
(branching) points---see, for example  \cite{kopp}. Consider now the
tessellation of the $z$--plane by means of  consecutive reflections of the domain
$ABC$ with respect to its sides, and  the corresponding reflections (inversions) of
the domain ${\cal ABC}$ in  the $\ze$--plane. Few first generations are shown in
Fig.\ref{fig:lacun}. The  obtained upper half--plane ${\rm Im} \ze >0$ has
a "lacunary" structure and represents the topological phase space of the
trajectories entangled with  the lattice of obstacles. The details of such
a construction as well as a  discussion of the topological features of the
conformal mapping $z(\ze)$ in the  symmetric case  can be found  in
\cite{nechaev}. We recall  the basic properties of the
transform $z(\ze)$ related to our  investigation of multifractality.  

The topological state of a trajectory $C$ in the lattice of obstacles can
be characterized as follows. 
\begin{itemize}
\item Perform the conformal mappings $z_{\rm s}(\ze)$ (or $z_{\rm ns}(\ze)$) of
the plane $z$ with symmetric (or nonsymmetric) triangular lattice of obstacles
to the upper half-plane ${\rm Im}\ze>0$,  playing the role of the topological phase
space of the  given model. 
\item Connect by nodes the centers of neighboring curvilinear triangles in the
upper half--plane ${\rm Im}\ze>0$ and raise a graph  $\gamma_{\rm s}$ (or 
$\gamma_{\rm ns}$) (which is, as shown below an isometric Cayley tree  embedded in the Poincar\'e plane).
\item Find the image of the path $C$ in the "covering space" ${\rm Im}\ze>0$
and define the shortest (primitive) path connecting the  centers of the
curvilinear triangles where the ends of the path $C$ are located. The
configuration of this primitive path projected to  the Cayley tree 
$\gamma_{\rm s}$ (or $\gamma_{\rm ns}$) plays the  role of topological invariant 
for the model under consideration. 
\end{itemize} 

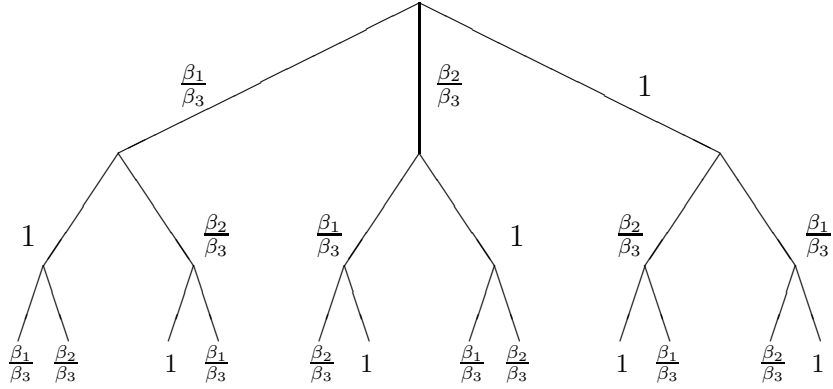
\begin{figure}
\begin{center}
\unitlength=1.00mm
\special{em:linewidth 0.6pt}
\linethickness{0.6pt}
\begin{picture}(113.33,55.00)
\put(20.00,35.00){\line(-2,-3){10.00}}
\put(20.00,35.00){\line(2,-3){10.00}}
\put(60.00,35.00){\line(-2,-3){10.00}}
\put(60.00,35.00){\line(2,-3){10.00}}
\put(100.00,35.00){\line(-2,-3){10.00}}
\put(100.00,35.00){\line(2,-3){10.00}}
\put(10.00,20.00){\line(-1,-3){3.33}}
\put(10.00,20.00){\line(1,-3){3.33}}
\put(30.00,20.00){\line(-1,-3){3.33}}
\put(30.00,20.00){\line(1,-3){3.33}}
\put(90.00,20.00){\line(-1,-3){3.33}}
\put(90.00,20.00){\line(1,-3){3.33}}
\put(110.00,20.00){\line(-1,-3){3.33}}
\put(110.00,20.00){\line(1,-3){3.33}}
\put(50.00,20.00){\line(-1,-3){3.33}}
\put(50.00,20.00){\line(1,-3){3.33}}
\put(70.00,20.00){\line(-1,-3){3.33}}
\put(70.00,20.00){\line(1,-3){3.33}}
\put(30.00,44.00){\makebox(0,0)[cc]{$\frac{\beta_1}{\beta_3}$}}
\put(64.00,44.00){\makebox(0,0)[cc]{$\frac{\beta_2}{\beta_3}$}}
\put(90.00,44.00){\makebox(0,0)[cc]{$1$}}
\put(8.00,24.00){\makebox(0,0)[cc]{$1$}}
\put(33.00,24.00){\makebox(0,0)[cc]{$\frac{\beta_2}{\beta_3}$}}
\put(48.00,24.00){\makebox(0,0)[cc]{$\frac{\beta_1}{\beta_3}$}}
\put(73.00,24.00){\makebox(0,0)[cc]{$1$}}
\put(88.00,24.00){\makebox(0,0)[cc]{$\frac{\beta_2}{\beta_3}$}}
\put(113.00,24.00){\makebox(0,0)[cc]{$\frac{\beta_1}{\beta_3}$}}
\put(7.00,7.00){\makebox(0,0)[cc]{{\foot $\frac{\beta_1}{\beta_3}$}}}
\put(13.00,7.00){\makebox(0,0)[cc]{{\foot $\frac{\beta_2}{\beta_3}$}}}
\put(27.00,7.00){\makebox(0,0)[cc]{{\foot $1$}}}
\put(33.00,7.00){\makebox(0,0)[cc]{{\foot $\frac{\beta_1}{\beta_3}$}}}
\put(47.00,7.00){\makebox(0,0)[cc]{{\foot $\frac{\beta_2}{\beta_3}$}}}
\put(53.00,7.00){\makebox(0,0)[cc]{{\foot $1$}}}
\put(67.00,7.00){\makebox(0,0)[cc]{{\foot $\frac{\beta_1}{\beta_3}$}}}
\put(73.00,7.00){\makebox(0,0)[cc]{{\foot $\frac{\beta_2}{\beta_3}$}}}
\put(87.00,7.00){\makebox(0,0)[cc]{{\foot $1$}}}
\put(93.00,7.00){\makebox(0,0)[cc]{{\foot $\frac{\beta_1}{\beta_3}$}}}
\put(107.00,7.00){\makebox(0,0)[cc]{{\foot $\frac{\beta_2}{\beta_3}$}}}
\put(113.00,7.00){\makebox(0,0)[cc]{{\foot $1$}}}
\put(60.00,55.00){\line(-2,-1){40.00}}
\put(60.00,55.00){\line(2,-1){40.00}}
\put(60.00,55.00){\line(0,-1){20.00}}
\end{picture}
\end{center}
\caption{Nonsymmetric 3--branching Cayley tree.}
\label{fig:3_cayley}
\end{figure}

The Cayley trees $\gamma_{\rm s,ns}$ have the same topological content as the 
one described in the  Section \ref{sect:2}, but here we  determine the Boltzmann weights
$\beta_1,\beta_2,\beta_3$ associated with passages between neighboring vertices
(see Fig.\ref{fig:3_cayley}) directly from the metric
properties  of the topological phase space obtained via the conformal
mappings $z_{\rm s,ns}(\ze)$.

It is well known that  random walks are  conformally invariant; in other words the
diffusion equation on the plane $z$ preserves  its structure under a
conformal transform,  but the diffusion coefficient can become space--dependent \cite{mckean}. Namely, under the conformal transform $z(\ze)$
the Laplace operator $\Delta_z=\frac{d^2}{dz d\overline{z}}$ is transformed in
the following way 
\be \label{laplace} 
\frac{d^2}{dz d\overline{z}}=
\frac{1}{|z'(\ze)|^2}\;\frac{d^2}{d\ze d\overline{\ze}} 
\ee

Before discussing the properties of the Jacobians $|z'(\ze)|^2$ for the 
symmetric and nonsymmetric transforms, it is more convenient to set up the 
following geometrical context. The connection between Cayley trees and 
surfaces of constant negative curvature has already been pointed out 
\cite{nechaev}, mostly through volume growth considerations. Therefore 
it becomes more natural to regard the  upper half--plane 
${\rm Im}\ze>0$ as the standard realization of the hyperbolic 2--space (surface of  constant negative curvature $R$, with here arbitrarily $R=-2$), that is to consider the following metric:
\be\label{met}
ds^2=\frac{-2}{R\eta^2}(d\xi^2+d\eta^2)
\ee
Let us  rewrite the Laplace operator (\ref{laplace}) in the form
\be \label{laplace2}
\frac{d^2}{dz d\overline{z}}=D(\xi,\eta)\;\eta^2\,
\left(\frac{d^2}{d\xi^2} + \frac{d^2}{d\eta^2}\right)
\ee
where the value $D(\xi,\eta)\equiv D(\ze)$ can be interpreted as the 
normalized space--dependent diffusion coefficient on the  Poincar\'e upper half--plane:
\be \label{diffusion}
D(\ze)=\frac{1}{\eta^2\,|z'(\ze)|^2}
\ee

The methods providing the conformal transform $z_{\rm s}(\ze)$ for the
symmetric triangle with angles $(\pi/3,\pi/3,\pi/3)$ have been discussed in
details in \cite{kopp}. The generalization of these results to the conformal
transform $z_{\rm ns}(\ze)$ for the nonsymmetric triangle with angles
$(\pi/2,\pi/6,\pi/3)$ is very straightforward. We here expose the Jacobians of those conformal mappings without derivation:
\be\label{jac}
\begin{array}{l}
\disp \left|z_{\rm s}'(\ze)\right|^2=\frac{1}{\pi^{2/3}B^2\left(\frac{1}{3},
\frac{1}{3}\right)}\;\left|\theta_{1}'(0,e^{i\pi\ze})\right|^{8/3} \medskip \\
\disp \left|z_{\rm ns}'(\ze)\right|^2=\frac{\pi^2}{B^2\left(\frac{1}{2},
\frac{1}{3}\right)}\left|\theta_{0}(0,e^{i\pi\ze})\right|^{8/3}\;
\left|\theta_{2}(0,e^{i\pi\ze})\right|^{4}\;
\left|\theta_{3}(0,e^{i\pi\ze})\right|^{4/3}
\end{array}
\ee
where $\theta_1'(\chi|...)=\frac{d}{d\chi}\theta_1(\chi|...)$ and
$\theta_i(0|...)$ $(i=0,...,3)$ are the standard definitions of Jacobi elliptic
functions \cite{abr}.

\begin{figure}[ht]
\begin{center}
\epsfig{file=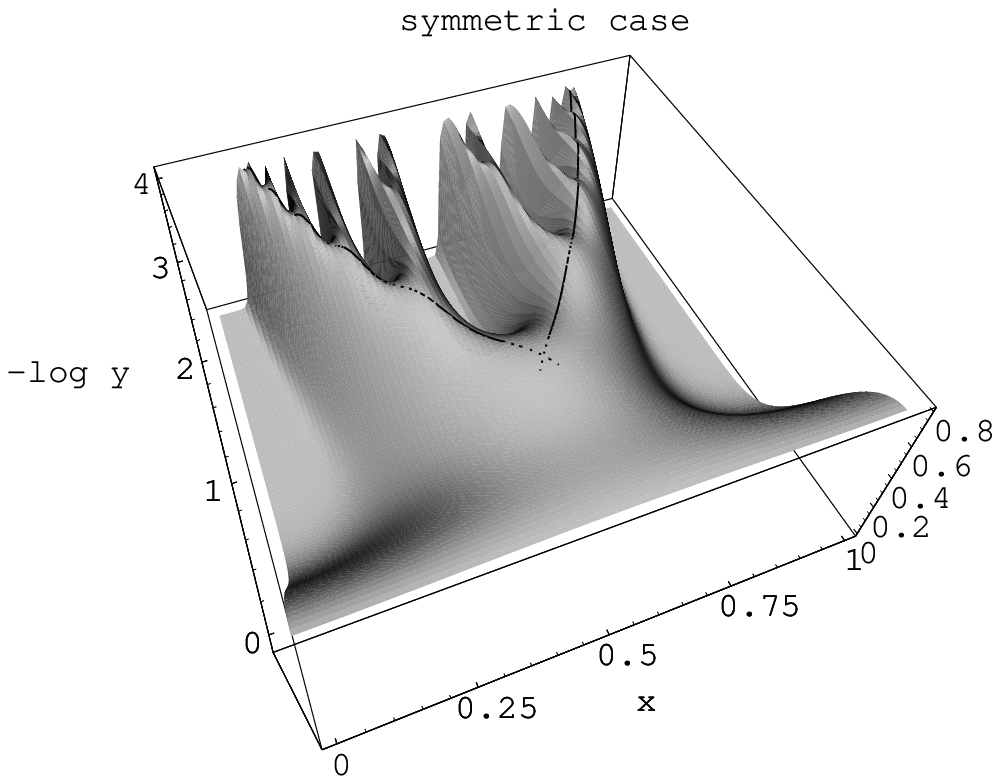,width=8cm}
\epsfig{file=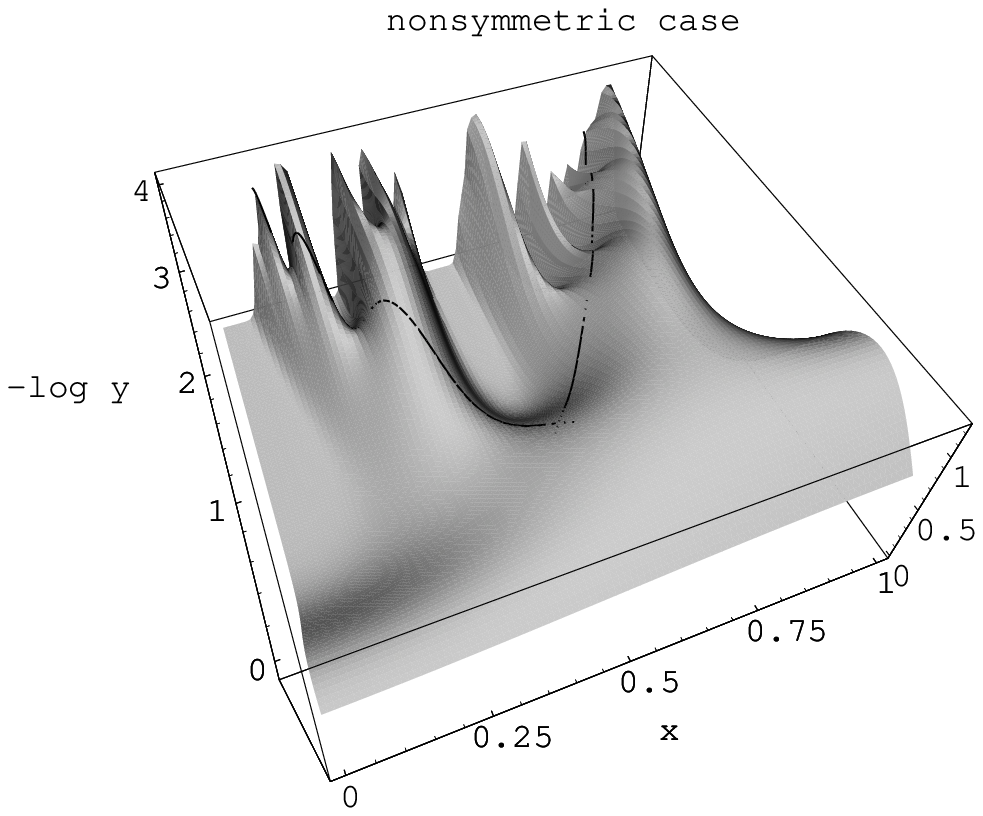,width=8cm}
\end{center}
\caption{Profiles of the surfaces $D^{-1}_{\rm s}(\xi,\eta)$ and
$D^{-1}_{\rm ns}(\xi,\eta)$ where $\xi\equiv x,\; \eta\equiv -\log y$. First 
generation of horocycles are shown by the lines.} 
\label{prof}
\end{figure}

Combining (\ref{diffusion}) and (\ref{jac}) we define the effective inverse 
diffusion coefficients in  symmetric ($D^{-1}_{\rm s}$) and in
nonsymmetric ($D^{-1}_{\rm ns}$) cases:
\be \label{diff_s_ns}
\begin{array}{l}
D^{-1}_{\rm s}(\ze)=\eta^2\;\left|z'_{\rm s}(\ze)\right|^2 \medskip \\
D^{-1}_{\rm ns}(\ze)=\eta^2\;\left|z'_{\rm ns}(\ze)\right|^2
\end{array}
\ee
The corresponding 3D plots of the reliefs $D^{-1}_{\rm s}(\xi,\eta)$ and
$D^{-1}_{\rm s}(\xi,\eta)$ are shown in Fig.\ref{prof}.

The functions $D^{-1}_{\rm s}(\zeta)$ and $D^{-1}_{\rm ns}(\zeta)$ are
considered as  quantitative indicators of the topological structure of the
phase spaces; in particular a Cayley tree can be isometrically embedded in the surface $D^{-1}_{\rm s}(\zeta)$. It can be  shown that the images of the centers of the triangles of the
symmetric lattice  in the $z$--plane correspond to the local maxima of the surface
$D^{-1}_{\rm  s}(\ze)$ in the $\ze$--plane. We define the vertices of the embedded tree as those maxima. The links connecting neighboring vertices are defined in the next paragraph.

Let us define the {\it horocycles} which correspond to repeating sequences of
weights in Fig.\ref{fig:3_cayley} with minimal periods. There are only three
such sequences: $\beta_1\beta_2\beta_1\beta_2...$, ~$\beta_1\beta_3
\beta_1\beta_3...$ and ~$\beta_2\beta_3\beta_2 \beta_3...$. The horocycles  are
images (analytically known) of certain  circles of the $z$--plane. They proved to be   a convenient tool for a constructive description of the
trajectories in  the $z$--plane starting from the trajectories in the covering space $\zeta$. 

The first generation of horocycles (closest to the root point of the Cayley 
tree) is shown in Fig.\ref{prof}. Let us consider the symmetric
case. Following a given horocycle  we follow a ridge of the
surface, and  we pass through certain maxima of this surface (that is through certain vertices of the tree). We therefore define {\it locally} the links of the tree as the set of ridges connecting neighboring maxima of  $D^{-1}_{\rm s}(\xi,\eta)$.  We recall that the ridge of the  surface can be defined as the set of points where the gradient of
the  function $D^{-1}_{\rm s}(\xi,\eta)$ is minimal along its isoline. Even if this gives a proper definition of the tree, extracting a direct parametrization is  difficult, that is why henceforth we will approximate the tree by arcs of horocycles.

To give a quantitative
formulation of the local definition of the embedded Cayley tree, we consider
the path  integral formulation of the problem on the $\zeta$--plane. Define
the  Lagrangian ${\cal L}\propto D^{-1}_{\rm s}(\ze){\dot \ze}^2$ of a free
particle  moving with the diffusion coefficient $D_{\rm s}(\zeta)$ in the  space
$\zeta$. Following  the canonical procedure and minimizing the corresponding
action  \cite{feynman}, we get the equations of motion in the effective
potential  $U=\ln(\eta^{2}D_{\rm s})$:
\be \label{euler}
\ddot q_{i}=({\dot q_j}\partial_{j}U){\dot q_i}-
\frac{1}{2}{\dot q_j}{\dot q_j}\partial_{i}U 
\ee
where $q_1=\xi$ and $q_2=\eta$. Even if Eq.(\ref{euler}) is nonlinear with a 
friction term, one can show that the trajectory of extremal action  between the centers of two neighboring triangles follow the ridge of
the surface $D_{\rm s}(\zeta)$. 

It is noteworthy that obtaining an analytical support of Cayley graphs is 
of great importance, since those graphs clearly display ultrametric 
properties and have  connections to $p$--adic surfaces \cite{freund}. 
The detailed study of metric properties of the functions $D^{-1}_{\rm s}(\ze)$ 
and $D^{-1}_{\rm ns}(\ze)$ is left for separate publication.

While the self--similar properties of the Jacobians of those conformal mappings
appear clearly in Fig.\ref{prof}, one  could wonder how the local symmetry breaking
affects the continuous problem. We  can see that if $D^{-1}_{\rm s}(\ze)$ is
univalued along the embedded tree,  $D^{-1}_{\rm ns}(\ze)$ does vary, what
makes the tree  locally nonuniform and leads to a  multifractal behavior. In other
words, different paths of same length along  the tree have the same weights in
the symmetric case, but have different ones  in the nonsymmetric case.  The
probability of a random path $C$ of length $L$ can be written in terms of a
path integral with a Wiener measure
\be \label{weight}
p_C={\cal D}\{s\}\exp\left\{-\int_0^{L}\frac{1}{D[s(t)]}
\left(\frac{ds}{dt}\right)^{2}dt\right\}
\ee
where $s(t)$ is a parametric representation of the path $C$.

\begin{figure}[ht]
\begin{center}
\epsfig{file=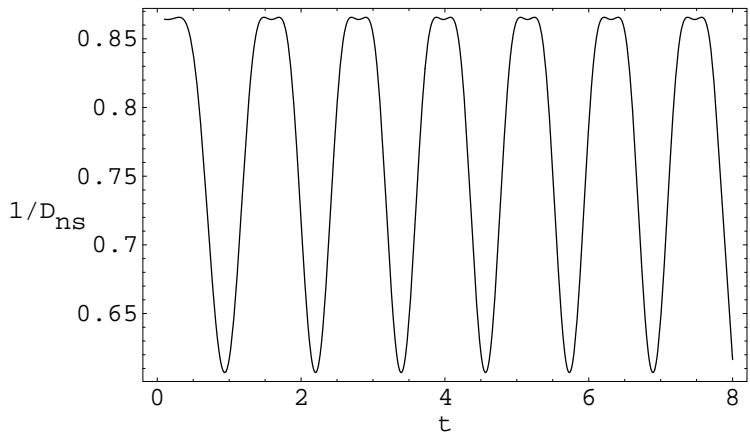,width=10cm}
\end{center}
\caption{$D^{-1}_{\rm ns}$ running along a horocycle at constant velocity.
Periodicity shows that the tree is isometric.}
\label{iso}
\end{figure}

The first horocycles in Fig.\ref{prof} can be parameterized as follows
\be\label{para}
\left\{\begin{array}{l}
\xi=\frac{1}{2}\pm (\frac{1}{2}-\frac{\sqrt{3}}{3}\sin\theta) \\
\eta=\frac{\sqrt{3}}{3}(1-\cos\theta)
\end{array}
\right.
\ee
with $\theta$ running in the interval $[0,\pi/2]$. The condition ensuring the
constant velocity $\dot{s}\equiv \frac{ds}{dt}$ along the horocycles gives 
with (\ref{met})
$$
\frac{1}{\eta}\; \frac{d\theta}{dt}={\rm const}
$$
hence
\be\label{t}
\theta(t)=\arctan\left(\frac{1}{t}\right)
\ee
with proper choice of the time unit. This parameterization is used to check 
that the embedded tree is  isometric. Indeed, the horocycles shown in 
Fig.\ref{prof} correspond to a periodic sequence of steps like 
$\beta_1\beta_2\beta_1\beta_2...$, $\beta_1\beta_3 \beta_1\beta_3...$ or 
$\beta_2\beta_3\beta_2\beta_3...$. It is natural to assert that a  step carries
a Boltzmann weight  characterized by the corresponding local values of  $D^{-1}_{\rm
ns}$. Therefore the period of the plot shown in Fig.\ref{iso}  is directly
linked to the spacing of the  tree embedded in the profile $D^{-1}_{\rm
ns}$.

\begin{figure}[ht]
\begin{center}
\epsfig{file=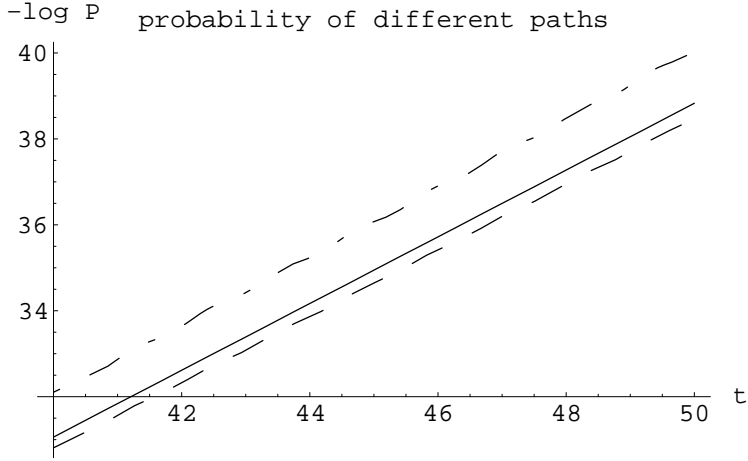,width=10cm}
\end{center}
\caption{Probability of different paths along horocycles. Dashed and 
dot--dashed  curves correspond to right-hand-- and left--hand--side
horocycles on  the nonsymmetric surface, while solid line shows both
right--hand-- and  left--hand--side horocycles on the symmetric surface.}
\label{slo}
\end{figure}

Coming back to the probability of different paths covered at constant 
velocity, one can write
\be \label{heights}
-\log p_C\propto\int_{t_1}^{t_2}\frac{dt}{D[s(t)]}
\ee
The figure \ref{slo} shows the value $-\log p_C$ in symmetric and
nonsymmetric cases for different paths starting at $t_1=0^{+}$ and ending at
$t$. In the symmetric case all plots are the same (solid line), whereas in
the nonsymmetric case they are different: dashed and dot--dashed curves
display the corresponding plots for the sequences $\beta_2\beta_3\beta_2
\beta_3...$  and $\beta_1\beta_3\beta_1 \beta_3...$.

Following the outline of construction of the fractal dimensions $D_q$ in Section \ref{sect:2},  we can describe multifractality in the continuous 
case by 
\be \label{haus} 
\disp D_q=-\frac{1}{q-1}\;\lim_{L\to\infty}\frac{1}{\ln{\cal N}(L)}
\ln\frac{\disp\int{\cal D}\{s\}\exp\left\{-q\int_0^L D^{-1}_{\rm ns}
(s(t))dt\right\}}{\disp \left[\int{\cal D}\{s\}\exp
\left\{-\int_0^L D^{-1}_{\rm ns}(s(t))dt\right\}\right]^q}
\ee
where ${\cal N}(l)$ is the area of the surface covered by the trajectories of
length $L$. This form is consistent with definitions \ref{1:Dq} and \ref{1:7}.  Indeed, if instead of the usual Wiener measure one chooses a 
discrete measure $d\chi_T$, which is nonzero only for trajectories along  the
Cayley tree,  we recover the following description. 

Define the distribution function $\Theta(\beta_1,\beta_2,\beta_3,k)\equiv 
\Theta\left(\frac{\beta_1}{\beta_3}, \frac{\beta_2}{\beta_3},k\right)$,  which has sense of the weighted number of directed paths of $k$ steps on the nonsymmetric  3--branching
Cayley tree shown in Fig.\ref{fig:3_cayley}. The values of the  effective Boltzmann
weights $\frac{\beta_1}{\beta_3}$ and  $\frac{\beta_2}{\beta_3}$ are defined in
terms of the local heights of  the surface $D_{\rm ns}^{-1}$ along the
corresponding branches of the  embedded tree. We set 
\be 
\begin{array}{l} 
\disp \frac{\beta_1}{\beta_3}= 
\exp\left[\disp \int_{t_1}^{t_2}\frac{dt}{D_{\rm ns}[s_{r}(t)]}- 
\int_{t_2}^{t_3}\frac{dt}{D_{\rm ns}[s_{r}(t)]}\right]\approx 1.07; \medskip \\ 
\disp \frac{\beta_2}{\beta_3}= 
\exp\left[\disp \int_{t_1}^{t_2}\frac{dt}{D_{\rm ns}[s_{l}(t)]}- 
\int_{t_2}^{t_3}\frac{dt}{D_{\rm ns}[s_{l}(t)]}\right] \approx 1.19 
\end{array} 
\ee
where $t_1,t_2,t_3$ are adjusted so that $s_{r}(t)$ represents a step weighted
with $\beta_3$ for $t_1<t<t_2$ and a step weighted with $\beta_1$ for
$t_2<t<t_3$ for right--hand--side horocycles while $s_{l}(t)$ represents a step
weighted with $\beta_3$ for $t_1<t<t_2$ and a step weighted with $\beta_2$ for
$t_2<t<t_3$ for left--hand--side horocycles.

The partition function $\Theta\left(\frac{\beta_1}{\beta_3},
\frac{\beta_2}{\beta_3},k\right)$ can be computed via straightforward
generalization of Eq.(\ref{1:6}); it can be written in the form:
\be
\Theta\left(\frac{\beta_1}{\beta_3},\frac{\beta_2}{\beta_3},k\right)=
A_0 \lambda_1^{k-1} + B_0 \lambda_2^{k-1} +  C_0 \lambda_3^{k-1} \qquad
(k\ge 1)
\ee
where
$\lambda_1$, $\lambda_2$ and $\lambda_3$ are the roots of the cubic
equation
$$
\lambda^3-\lambda\left(1+\frac{\beta_2^2}{\beta_3^2}+
\frac{\beta_1\,\beta_2}{\beta_3^2}\right)-
\left(\frac{\beta_1\,\beta_2}{\beta_3^2}+
\frac{\beta_2^2}{\beta_3^2}\right)=0
$$
and $A_0$,$B_0$ and $C_0$ are the solutions of the following system of linear
equations
$$
\left\{\begin{array}{l}
\disp A_0+B_0+C_0=1+\frac{\beta_1}{\beta_3}+\frac{\beta_2}{\beta_3}
\medskip \\
\disp A_0 \lambda_1 + B_0 \lambda_2 +  C_0 \lambda_3=
2\frac{\beta_1}{\beta_3}+2\frac{\beta_2}{\beta_3}+
2\frac{\beta_1\beta_2}{\beta_3^2}
\medskip \\
\disp A_0 \lambda_1^2 + B_0 \lambda_2^2 +  C_0 \lambda_3^2=
\frac{\beta_1}{\beta_3}+\frac{\beta_2}{\beta_3}+
\frac{\beta_2^2}{\beta_3^2}+\frac{\beta_1^2}{\beta_3^2}+
6\frac{\beta_1\beta_2}{\beta_3^2}+\frac{\beta_1^2\beta_2}{\beta_3^3}+
\frac{\beta_1\beta_2^2}{\beta_3^3}
\end{array}\right.
$$

\begin{figure}
\centerline{\epsfig{file=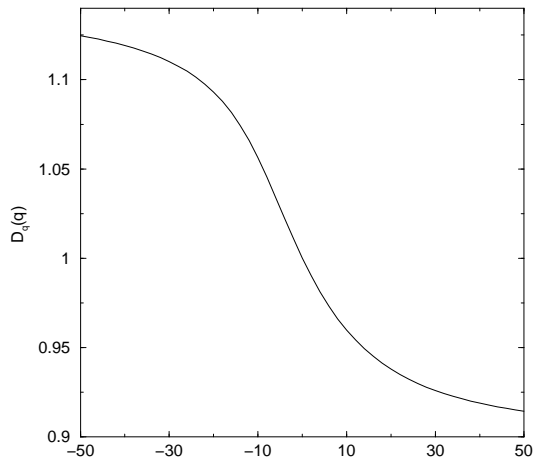,width=7cm}}
\caption{Multifractality of trajectories on nonsymmetric tree with Boltzmann
weights defined by the Jacobian of the conformal mapping (\ref{diff_s_ns}).}
\label{fig:haus2}
\end{figure}

Knowing the distribution function $\Theta\left(\frac{\beta_1}{\beta_3},
\frac{\beta_2}{\beta_3},k\right)$, Eq.(\ref{haus}) with the discrete measure 
$d\chi_T$ reads now (compare to (\ref{1:7})--(\ref{1:7a}))
\be \label{haus2}
D_q=-\frac{1}{q-1}\lim_{k\to\infty}\frac{\ln
\Theta\left(\left[\frac{\beta_1}{\beta_3}\right]^q,
\left[\frac{\beta_2}{\beta_3}\right]^q,k\right)-
q\ln \Theta\left(\frac{\beta_1}{\beta_3},
\frac{\beta_2}{\beta_3},k\right)}{\ln (3 \times 2^{k-1})}
\ee
The plot of the function $D_q(q)$ is shown in Fig.\ref{fig:haus2} (the 
plot is drawn for $k=100\,000$).

\section{Discussion}
The results presented  in Sections \ref{sect:2}--\ref{sect:4} are summarized; they underline several problems still unsolved related to our work,  and raise the issue of their possible applications to real physical systems.

1. The basic concepts of multifractality have been clearly formulated mainly
for abstract systems in \cite{procacc}. In the present work, we have tried to remain
as close as possible to these classical formulations, while adding to  abstract
models of Ref.\cite{procacc}  the new physical content of topological  properties of random walks entangled with an array of obstacles. Our results point out two conditions which generate multifractality for any  physical system: (i) an exponentially growing number of  states, i.e.
"hyperbolicity" of the phase space, and (ii) the breaking of a  local symmetry of
the phase space (while on large scales the phase space could remain isotropic). 

In Section \ref{sect:2} we have considered the topological properties of 
the discrete  "random walk in a  rectangular lattice of  obstacles" model.
Generalizing an approach developed earlier (see for example \cite{nechaev} and
references  therein) we have shown that  the topological phase space of the model is a Cayley tree whose associated transition probabilities are nonsymmetric. Transition probabilities have been computed from the  basic characteristics of a free random walk within the elementary cell of the lattice of obstacles. The family of generalized Hausdorff dimensions
$D_q(q)$ for the partition function $\Omega(\beta^q,k)$ (where $k$ is the
distance on the Cayley graph which  parameterizes the topological state of the
trajectory) exhibits nontrivial dependence on $q$, what means that different moments of
the partition function $\Omega(\beta^q,k)$ scale in different ways, e.g.  that $\Omega(\beta^q,k)$ is multifractal.

The main topologically--probabilistic issues concerning the distribution of
 random walks in a rectangular lattice of obstacles have been considered
in  Section \ref{sect:3}. In particular we have computed the average "degree of
entanglement" of a $\tilde{N}$--step random walk and the probability for a $\tilde{N}$--step random
walk to be closed and unentangled. Results have been achieved through a renormalization group technique 
on a nonsymmetric Cayley tree. The renormalization procedure has allowed us to
overcome one major  difficulty: in spite of  a locally broken spherical symmetry, we have mapped our problem to a symmetric random walk on a tree of effective branching number $z$ depending on the lattice parameters. To validate  our procedure, we have compared the return
probabilities obtained via our RG--approach with  the exact result of P.Gerl and
W.Woess \cite{woess} and found a very good numerical agreement.  

The problem tackled in Section \ref{sect:4} is closely related to the one 
discussed in Section \ref{sect:2}. We believe that the 
approach developed in Section \ref{sect:4} could be very important and
informative as  it explicitly shows that  multifractality is not
attached to particular properties of a statistical system (like random walks in our case) but deals directly with  
metric properties of the   topological phase space. As we have already pointed out,  the required conformal transforms are known only for triangular lattices, what restricts our study. However we explicitly showed that the transform $z_{\rm ns}(\ze)$ maps the multi-punctured complex plane $z$ onto the so-called ``topological phase space'', which is the complex plane $\ze$ free of topological obstacles (all obstacles are mapped onto the real axis). We have connected multifractality to the multi--valley structure of the properly normalized
Jacobian $D_{\rm ns}(\xi,\eta)$ of the nonsymmetric conformal mapping
$z_{ns}(\ze)$. The conformal mapping obtained has deep relations with number theory, which  we are going to discuss in a  forthcoming publication.

2. The "Random Walk in an Array of Obstacles"--model can be considered as a
basis of a mean--field--like approach to the problem of
entropy calculations in sets of strongly entangled fluctuating polymer
chains. Namely, we choose a  test chain, specify its topological state and assume
that the lattice of obstacles models the effect of entanglements with the
surrounding chains (the "background"). Changing $c_x$ and $c_y$ one can mimic
the affine deformation of the background. Investigating the free energy
of the test chain entangled with the deformed media is an important step towards understanding high-elasticity of polymeric rubbers \cite{khter}.

Neglecting the fluctuations of the background as well as the topological
constraints which the test chain produces by itself, leads to  information
 losses about the correlations between the test chain and the background. Yet, even in this simplest case we  obtain  nontrivial statistical results concerning the test chain topologically  interacting with the
background.

The first attempts to go beyond the mean--field approximation of
RWAO--model and to develop a  microscopic approach to statistics of mutually
entangled chain--like objects have been undertaken recently in
\cite{des_ver_never}. We believe that investigating multifractality of such
systems is worth attention.
\bigskip

{\bf Acknowledgments}
\medskip

The authors are grateful to A.Comtet for valuable discussions and helpful
comments, and would like to thank the referees for drawing their attention to
references  \cite{lyons,pem,holley}.

\end{document}